\documentclass[11pt,a4paper]{emulateapj}
\usepackage{natbib}
\usepackage{amssymb, amsmath, amsbsy}
\usepackage{color}
\usepackage{rotating} 
\usepackage[linkcolor=blue]{hyperref}%

\usepackage{epsfig}
\usepackage{amsmath}
\usepackage{amssymb}
\usepackage{natbib}
\usepackage{subfigure}
\usepackage{graphicx}

\slugcomment{Accepted to AAS Journals}

\newcommand{\civ}{C {\sc iv}}
\newcommand{\heii}{He {\sc ii}}

\newcommand{\mgii}{Mg {\sc ii}}

\begin{document}
\def\mean#1{\left< #1 \right>}
\newcommand{\sbunit} {
  erg\ s$^{-1}$ cm$^{-2}$ arcsec$^{-2}$}
  
\title{Evolution of the Cool Gas in the Circumgalactic Medium (CGM) of Massive Halos -- A Keck Cosmic Web Imager (KCWI) Survey of Ly$\alpha$ Emission around QSOs at $z\approx2$}
\author{Zheng Cai$^{1,2}$; Sebastiano Cantalupo$^3$; 
J. Xavier Prochaska$^{2,4}$; Fabrizio Arrigoni Battaia$^5$;
Joe Burchett$^2$; Qiong Li$^6$; John Chisholm$^2$; 
Kevin Bundy$^2$; 
Joseph F. Hennawi$^7$ }
\affil{$^1$  Department of Astronomy, Tsinghua University, Beijing 100084, China}
\affil{$^2$ UCO/Lick Observatory, University of California, 1156 High Street, Santa Cruz, CA 95064, USA }
\affil{$^3$ Department of Physics, ETH Zurich, Wolfgang-Pauli-Strasse 27, CH-8093 Zurich, Switzerland}
\affil{$^4$ Kavli Institute for the Physics and Mathematics of the Universe (WPI), The University of Tokyo, Kashiwa 277-8583, Japan}
\affil{$^5$ Max-Planck Institut fur Astrophysik, Karl-Schwarzschild-Strasse 1, 85748 Garching, Germany}
\affil{$^6$ Kavli Institute for Astronomy and Astrophysics, Peking University, Beijing 100871, China}
\affil{$^7$ Department of Physics, Broida Hall, University of California at Santa Barbara, Santa Barbara, CA 93106, USA}



 

\begin{abstract}
Motivated by the recent discovery of the near-ubiquity 
of Ly$\alpha$ emission around $z \gtrsim 3$ QSOs, 
we performed a systematic study of QSO circumgalactic 
Ly$\alpha$ emission at $z\approx2$, utilizing the 
unique capability of the Keck Cosmic Web Imager (KCWI) 
-- a new wide-field, blue sensitive integral-field 
spectrograph (IFU). In this paper, we present KCWI 
observations on a sample of 16 ultraluminous Type-I 
QSOs at $z=2.1-2.3$ with ionizing luminosities of 
$L_{\rm{\nu_{\rm{LL}}}}=10^{31.1-32.3}$ erg s$^{-1}$ Hz$^{-1}$. 
We found 
that { 14 out of 16} QSOs are associated with Ly$\alpha$ nebulae with projected linear-sizes larger 
than 50 physical kpc (pkpc). 
Among them, four nebulae have enormous Ly$\alpha$ emission with 
the Ly$\alpha$ surface brightness $SB_{\rm{Ly\alpha}} >10^{-17}$ erg s$^{-1}$ cm$^{-2}$ arcsec$^{-2}$ 
on the $>100$ kpc scale, 
extending beyond the field of view of KCWI. 
Our KCWI observations reveal that most $z\approx2$ QSO nebulae 
have a more irregular morphology compared to those at $z\gtrsim3$. 
In turn, we measure that the 
circularly-averaged surface brightness (SB) at $z\approx2$ 
is 0.4 dex fainter than the redshift-corrected,
median SB at $z\gtrsim 3$. The Ly$\alpha$ SB profile (SB$_{\rm{Ly\alpha}}$) of 
QSOs at $z\approx 2$ 
can be described by a power law of SB$_{\rm{Ly\alpha},z\approx2.3}= 
3.7\times10^{-17}\times(r/40)^{-1.8}$ erg s$^{-1}$ cm$^{-2}$ arcsec$^{-2}$, 
with the slope similar to that at $z\gtrsim3$. 
The observed lower redshift-corrected, circularly-averaged SB 
may be mainly due to the lower covering factor of cool gas clouds in massive halos 
at $z\approx2$. 
\end{abstract}

\keywords{Intergalactic medium, QSOs: emission lines, galaxy: halos} 



\section{Introduction}\label{sec:introduction}

Over the past decade, theoretical studies have established a new paradigm for 
the accretion of gas into dark matter halos 
to fuel star-formation during
galaxy formation \cite[e.g.,][]{dekel09, keres09, nelson13, vogelsberger13}. 
This model predicts that galaxies are
fed by cool `streams' of gas, linked to the surrounding intergalactic medium 
(IGM) by a web of cosmic filaments \cite[e.g.,][]{bond96, fukugita98}. These 
filaments contain a rich reservoir of nearly pristine gas that drives galaxy formation 
and evolution, especially in the early Universe \cite[e.g.,][]{keres05, dekel09,vandevoort11,fumagalli11, dekel13, correa15}.  
Nevertheless, this fundamental picture is difficult to test observationally.  
Direct imaging of the IGM is crucial for examining this standard paradigm of 
galaxy formation and further revealing the IGM-galaxy interactions. 

The detection of the IGM in emission was suggested a few 
decades ago \cite[e.g.,][]{hogan87,gould96}. 
Nevertheless, progress was hindered by the faintness of the IGM emission. 
By searching around luminous QSOs, the 
expected diffuse emission due to recombination should be 
enhanced by a few orders of magnitude 
within the densest part of the cosmic web \cite[e.g.,][]{cantalupo05, cantalupo12, kollmeier10}. 
 With narrowband 
imaging, \citet{cantalupo14, hennawi15} and \citet{cai17} discovered a few sources that are 
sufficiently luminous for quantitative analysis of diffuse gas emission. 
These sources define the ``Enormous Ly$\alpha$ Nebulae 
(ELANe)" \cite[e.g.,][]{cai17b, arrigoni18, cai18, arrigoni19}; 
they are the extrema of Ly$\alpha$ nebulosities at $z \sim 2-3$, with sizes exceeding the diameters of 
massive dark matter halos ($\sim 200$ kpc)  with SB$_{\rm{Ly\alpha}}\ge10^{-17}$ erg s$^{-1}$ cm$^{-2}$ 
arcsec$^{-2}$ and  
Ly$\alpha$ luminosities greater than 10$^{44}$ erg s$^{-1}$. 

Recent progress in wide-field 
integral-field spectrographs (IFS) 
on 8-10m telescopes, including MUSE 
and KCWI, provide us an indispensible opportunity to 
directly study the IGM/CGM around bright sources 
at $z=2-4$ by reaching an unprecedented, low surface brightness (SB) of 
 a few $\times 10^{-19}$\,erg s$^{-1}$ cm$^{-2}$ arcsec$^{-2}$. This 
makes it possible to study emission from the 
circumgalactic medium (CGM) and IGM around bright sources.  
{ With VLT/MUSE and using a sample of 17 QSOs}, \citet{borisova16} 
reveal that Ly$\alpha$ nebulae with projected sizes exceeding 100 kpc 
are ubiquitous for QSOs at $z \gtrsim 3$. \citet{arrigoni19} further confirmed such near-ubiquity 
using a sample of 61 QSOs at $z\gtrsim3$. 
Nevertheless, a systematic IFS survey has not been performed at $z<3$. 
A crucial question is whether one can construct a uniform 
sample at a lower redshift (e.g., at $z\sim2$), and probe evolution
in the IGM/CGM across cosmic time. 

At $z\approx 2$, \citet{arrigoni16} have conducted deep narrowband 
images on 15 $z\approx2.2$ QSOs which have a fainter luminosity 
compared with QSO sample at $z\gtrsim3$ \citep{borisova16, arrigoni19}.
 \citet{arrigoni16} did not detect bright nebulae with 
 the projected size of $>50$ kpc at Ly$\alpha$ SB of 10$^{-17}$ \sbunit.  
 Compared with the ubiquitous Ly$\alpha$ nebulae at $z\approx3$, the narrowband 
 results seem to suggest a strong evolution of Ly$\alpha$ emitting cool gas from 
 $z=3$ to $z=2$. However, before drawing such a conclusion, we note that the 
 following effects may partially yield such a low detection rate 
 of Ly$\alpha$ nebulae at $z\approx 2$ (also see \S4.1.3 in Arrigoni Battaia et al. 2019): 
(1) The redshift of nebular Ly$\alpha$ may have an offset with the systemic redshift 
determined by the \mgii\ emission. 
If we use a narrowband with the central wavelength consistent 
with the \mgii-determined systemic redshift, then the Ly$\alpha$ emission could reside 
outside the narrowband;
(2) The full-width-half-maximum (FWHM) of the 
narrowband is still much wider than the FWHM of the nebular Ly$\alpha$ emission. 
Thus, the contrast between the point-spread-function (PSF) and the diffuse Ly$\alpha$ 
emission is higher for the narrowband data comparing to the IFU data, which makes 
the PSF subtraction more difficult  for the narrowband data. 
{ (3) The surface brightness limit that MUSE reaches is deeper than the 
narrowband study which may also yield a higher detection rate of the Ly$\alpha$ 
nebulae.}

To build a uniform sample for studying IGM emission at $z\approx2$ and directly test 
its evolution, 
we have conducted a new survey using the Keck Cosmic Web Imager (KCWI) 
to search for Ly$\alpha$ nebulae associated with QSOs at $z\approx 2$. 
With the KCWI, we can conduct, for the first time, a fair comparison with $z\gtrsim3$ MUSE results, 
 by selecting QSOs with similar ionizing luminosities as that of the MUSE samples (Borisova et al. 2016; Arrigoni Battaia et al. 2019). 
Our goals are to understand the cool gas budget around 
 ultraluminous QSOs at $z\approx2$,  
 to study the kinematics throughout QSO halos, and to
 probe the evolution of the CGM around massive halos from 
$z=3-2$. 

In this paper, we present the first KCWI observations on the 
Ly$\alpha$ nebulae associated with ultralumious Type-I QSOs 
at $z=2.1-2.3$. We organize this paper as follows. In \S2, we introduce our observations and data reduction. 
In \S3, we provide the results of the KCWI observations, perform 
optimal extraction of Ly$\alpha$ emission around QSOs, 
and study the gas kinematics and morphology.  
In \S4, we provide a discussion of the results. The discussion is 
based on the observational results from 
the KCWI, MUSE, and previous narrowband studies. 
Throughout this paper when measuring distances, we refer to physical distances unless
otherwise specified.
We assume a $\rm{\Lambda}$CDM cosmology with 
$\Omega_m= 0.3$, $\Omega_{\Lambda}=0.7$ and $h=0.70$.

\section{Observations}\label{sec:observations}

In this section, we provide details on the 
KCWI instrument configurations, observations, 
data reduction pipeline, and post-processing 
after the standard reduction pipeline. 



\subsection{KCWI Instrument Configuration}

{Keck Cosmic Web Imager (KCWI) (e.g., Morrissey et al. 2018) is 
a general purpose, optical IFS that has been installed on the 10 m Keck II telescope. 
KCWI provides seeing-limited imaging from the wavelength range of 3500 -- 5700 \AA, 
and the spectral resolution can be configured from $R=1000$ to $R =20000$. The 
field of view  is $20''\times33''$ for large slicer, $16''\times20''$ for the medium slicer 
and $8''\times20''$ for the small slicer. 
KCWI is optimal for a survey of gaseous nebulae
at $z\approx2$ because: (1) KCWI has 
a high throughput from $\lambda = 3800 - 5500$\AA, optimal 
for probing the Ly$\alpha$, \civ, and \heii\ lines at $z\sim2$. 
(2) KCWI has high spectral-resolution modes ($R>4000$) 
which resolve the gas kinematics, 
and (3) KCWI has a relatively large 
field-of-view (FoV) to  cover extended Ly$\alpha$ nebulae.
Furthermore, KCWI nicely complements the characteristics
of MUSE which thrives at $\lambda > 5000$\AA.  }

Data was taken with the Keck/KCWI instrument between 
November 15 2017 and January 30 2019. The seeing 
varied in the range of 0.7 -- 1.1 arcsec (FWHM of the 
Gaussian at $\approx4000$\AA, measured in the combined 
40 min datacubes). The information of the QSO fields 
are summarized in Table 1.

For our program, we configured KCWI
with the BM grating and medium slicer which 
yields a FoV of 16.8$''$ {perpendicular to slicer} 
(24 slicers) and  20$''$ { along the slicer.} { 
We also use BM grating and large slicer to observe one of our QSOs: Q1444}

This FoV is sufficient to map the gas around QSO host halos
to a radius of $\approx 100$\,kpc at $z\approx2.3$. {This setting can provide} a spatial sampling
of { $ \frac{20''}{24\ {\rm{slicers}}}\approx 0.67''$} along the slicer and is seeing-limited 
perpendicular to the slicer. The spectral resolution 
is $R=4000$.  We observed at central wavelengths ranging 
from $\lambda= 3900$\AA\ -- 4100\AA\ to cover the Ly$\alpha$ emission of 
each QSO in the sample. 

The total exposure time for each target is 40 minutes, 
which consists of four 
10-minute individual exposures. The observing procedure is as follows: 
we divide the entire QSO sample into { several sub-groups}, 
with each sub-group consisting of two to several QSOs
separated by $\lesssim 3$ degrees on the sky. 
Also, we require that
the redshift offset $\Delta z$ between each of our QSOs is greater than 
$|\Delta z| > 0.05$. This procedure 
was taken because the sky is determined using a nearby  
offset-target with Ly$\alpha$ emission lines at 
different wavelengths (see details in \S2.3.3). 

\subsection{KCWI QSO Sample at $z\approx2$}

Our ultraluminous QSO sample at $z\approx2$ is selected from the 
SDSS-IV/eBOSS database \cite[e.g.,][]{paris17}, restricted to $2.1<z<2.3$. The lower 
limit of $z=2.1$ is set by the blue limit to the sensitivity of KCWI. 
The purpose for constraining $z\le 2.3$ is that there 
is currently no systematic IFU survey at $z\le2.3$.  
Further, at this redshift, H$\alpha$ emission can be observed 
from the ground for additional insight into the mechanism(s)
powering Ly$\alpha$ \citep[e.g.][]{leibler18}. 
Our KCWI sample can be compared with previous VLT/MUSE observations 
at $z>3$ \citep{borisova16,arrigoni19} and Gemini and Keck narrowband 
surveys at $z\approx2.3$ \cite[e.g.,][]{arrigoni16}. 

\begin{figure}[h]
\includegraphics[width=0.5\textwidth,height=0.28\textheight]{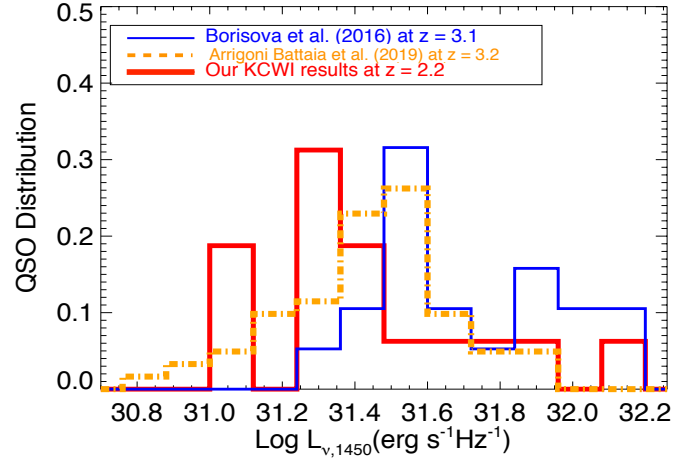}
 \caption{The distribution of the QSO luminosity at the 
 rest-frame of 1450 \AA\ at $z\approx2$ (Red line). 
 The $z\approx2$ QSO sample has a luminosity of  $L_{\nu, \rm{1450\AA}}= 10^{31.48\pm0.32}$ erg s$^{-1}$ Hz$^{-1}$. 
 The blue line represent the QSOs at $z\approx3$ from 
 Borisova et al. (2016) which { have} a QSO luminosity of 
 $L_{\nu,\rm{1450\AA}}= 10^{31.77\pm0.25}$ erg s$^{-1}$ Hz$^{-1}$. 
 The yellow dot-dashed line represents 
 the QSO luminosity at 1450 \AA\ at $z\approx3$ in 
 Arrigoni Battaia et al. (2019) which has a QSO luminosity of 
  $L_{\nu,\rm{1450\AA}}= 10^{31.49\pm0.23}$ erg s$^{-1}$ Hz$^{-1}$. 
 Our KCWI QSO sample has a similar {median bolometric} luminosity as 
 Arrigoni Battaia et al. (2019) and is $0.29$ dex fainter 
 than Borisova et al. (2016). } 
 \label{fig:sample}
\end{figure}

We select our QSO sample using the following criteria: (1) 
ultraluminous QSOs with $i_{\rm{mag}}<18.5$; 
(2) each source must have at
least one other ultraluminous QSOs within $3$\,degree separation, 
and with a redshift offset $|\Delta z| >0.05$ (see \S2.3.3 for 
details of sky subtraction).
Using these selection criteria, we  
select 16 QSOs (listed in Table~\ref{table:QSO_snapshot}). 
Note that the SDSS QSO 
density peak is at $z\approx2.3$ \cite[e.g.,][]{paris17}, providing us a large database 
for selecting targets. The QSOs selected have a median 
$i$-band magnitude of 17.7, 0.2 magnitude brighter than the 
median $i$-band magnitude of 17 $z\approx3.1$ QSOs in 
\citet{borisova16}, and 0.5 magnitude brighter than 
the 61 $z\approx3.2$ QSOs 
described in \citet{arrigoni19}. We further calculated 
$L_{\rm{1450}}$ which is the luminosity $\nu L_{\rm{\nu}}$ at 
$\lambda= 1450$\AA. For our KCWI QSO sample, the median 
$L_{\rm{1450}}$ is $1.7\times10^{13}$ $L_\odot$. The luminosity of our $z=2$ QSOs, 
on average, is similar 
to \cite{arrigoni19}.
The QSOs of \cite{borisova16} have a median 
luminosity of 
$3.2\times10^{13}$ $L_\odot$, i.e.\ $1.9\times$ 
the median $L_{\rm{1450}}$ of our KCWI QSO sample. 
We summarize the QSO luminosity of the 
three samples in Figure~\ref{fig:sample}. 

\subsection{Data Reduction}
\label{sec:data_reduction}

In this section, we provide a detailed description of the 
data reduction, including the standard KCWI pipeline  
and our post-processing steps using 
the CubeExtractor package
(Cantalupo et al. 2019). 

\subsubsection{KCWI Standard Pipeline }

The KCWI pipeline v1.0\footnote{ https://github.com/kcwidev/kderp/}, released in Mar. 17, 2018 was adopted to reduce 
our data \citep{morrissey18, cai18}. 
For each image, we first subtracted the bias, correcting the pixel-to-pixel 
variation using flat-field images, 
removing cosmic-rays, and error image creation. 
Then, continuum-bar images were used to conduct a
geometric transformation, and the ThAr arc images were analyzed for 
wavelength calibration. 
At this stage, the datacube was constructed. 
Then, the twilight flats were used to correct the 
slice-to-slice variance. Each individual image was 
  flux calibrated using a spectrophotometric standard star 
taken at the beginning of the night. 

\subsubsection{CubeFix: Improving the Flat-fielding of the Cubes}

Our scientific goals require the analysis of
diffuse, extended emission at low surface brightness. 
Therefore, we must control for systematic 
variations across the datacube. 
We used custom tools for flat-fielding correction 
and sky-subtraction. The procedures 
are part of the CubExtractor package \citep{cantalupo19} 
which was developed 
to improve the detection of faint, low surface brightness emission 
in IFU datacubes \citep[e.g. MUSE,][]{borisova16}. 
We used the CubeFix routine to correct the systematic 
errors due to flat-fielding. Then,  
we constructed a medium-band image which is collapsed 
using 300 channels in the wavelength direction. The 
slice-by-slice correction is then calculated using the medium-band image. 
The flat-fielding correction is performed as a self-calibration 
using sky-lines and sky continuum as a 
uniform source to re-calibrate each individual slice of the IFU. 
Sources are masked 
in this procedure to minimize the self-calibration errors. 
With these steps, the 
residual is at a level of less than 0.1\% of the sky.

\subsubsection{Sky Subtraction and Coaddition}

Sky-subtraction was then performed on each individual, flat-field corrected cube 
using our custom procedure (CubeSharp routine in the CubExtractor package). 
As noted above, we have associated each QSO with another that lies
within three degree separation on the sky. 
We estimate the sky, channel-by-channel, from the unsubtracted
datacube of the offset-target.
The sky is calculated by taking the median after masking sources. 
In each wavelength, all pixels with 3-$\sigma$ above the median {were} masked as sources, and we repeated this process 10 times to construct a final source mask. 
The offset-target always has a redshift offset of $\Delta z > 0.05$ from the main target 
({$\Delta z>0.05$ corresponds to $\gtrsim 120$ wavelength channels}). 
This criterion insures that 
the Ly$\alpha$ emission of the offset-target lies far from the emission from the target so that 
we can use the sky determined from the offset-target field around the wavelength of target's Ly$\alpha$ emission. 
For each exposure, we determined the QSO center; and then 
aligned each exposures according 
to this value. We 
performed a weighted mean with inverse variance
weighting to construct the final datacube.
The variance images we used are generated from the KCWI pipeline. 
{ The seeing condition for each target has been taken into account 
in each variance images. The variance amplitude is proportional to the inverse of the
 seeing full-width-half-maximum (FWHM).}


\subsection{Point-spread-function (PSF) and Continuum subtraction}

To search for extended, faint Ly$\alpha$ emission distinct
from the QSO flux, 
we applied the following procedure to subtract
the central, QSO emission.
For each wavelength channel, we produce a pseudo-narrowband image 
with a width of 300 spectral channels ($\approx150$ \AA).  
 For our analysis, we used the median of these $\approx300$ 
wavelength channels for constructing the pseudo-broadband images. We found that the number of wavelength 
channels we adopted provides a good compromise 
between capturing wavelength PSF variations and 
obtaining a good signal-to-noise ratio in the 
pseudo-broadband image. 
Then, we rescaled the empirical PSF 
according to the integrated flux within the $1''\times1''$ 
area around the QSO centroid. 
Then, we subtracted the PSF at each wavelength 
channel \cite[e.g.,][]{herenz15,borisova16}. 
We computed the rescaling factor using an averaged-sigma-clip algorithm to 
minimize cosmic ray effects. 



 We further removed continuum sources in each spaxel (a spectrum in the datacube) 
 of the cube using a median-filtering approach to construct 
 a continuum image. The window size 
 of the median filter is approximately 
 300 pixels ($\approx 150$ \AA).   We then subtract the 
 continuum image from each wavelength channel to construct a continuum-subtracted 
datacube. Sources with flat continua are expected to 
be eliminated with this procedure. 
 In several cases, stars or background galaxies 
 are not completely removed or 
 are over-subtracted by this procedure due to the large window size.  
  However, these residuals do not affect our results 
  because we mask stars 
  or galaxies with bright continuum in the cube before extraction. 
  
\begin{figure}[h]
\includegraphics[width=0.5\textwidth,height=0.29\textheight]{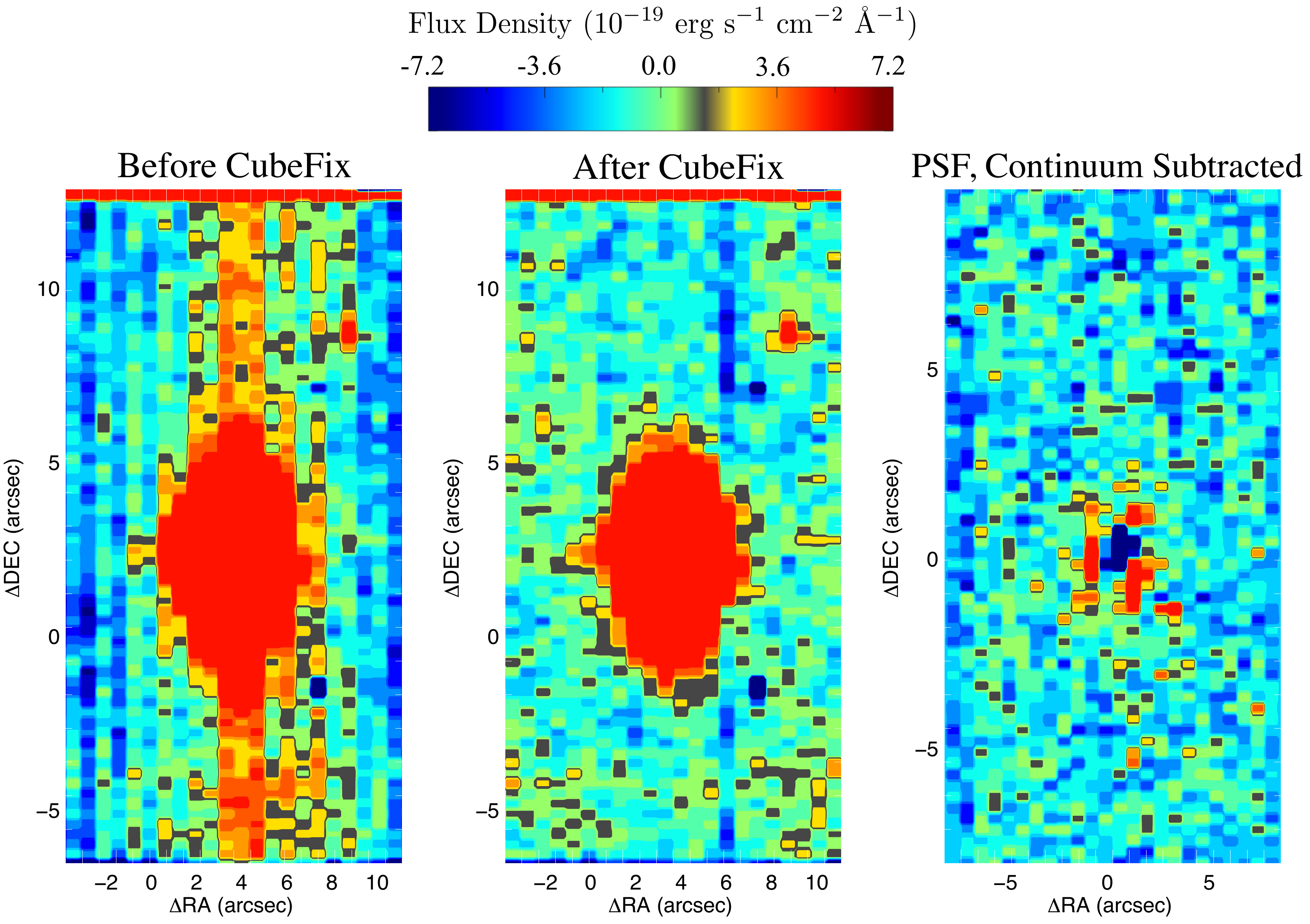}
 \caption{Left panel shows a white-light image obtained from a cube reduced with the KCWI standard pipeline 
 (v1.0).  Middle cube indicates 
 the data product after running our post-process script CubeExtractor (CubeFix routine). 
 Note that the scattered light is reduced. 
 Right cube shows the Point-spread-function (PSF) and continuum-subtracted 
 cube. The images shown here are constructed  using the median of 2000 km s$^{-1}$ 
 velocity channels between the Ly$\alpha$ and NV region, where we expect no significant line 
 emission in this velocity range, and the median flux is consistent with zero. } 
  \label{fig:reduction}
\end{figure}

  In the left panel of Figure~\ref{fig:reduction}, we show the data 
  product just after processing by the KCWI standard pipeline. 
  In the right panel of Figure~\ref{fig:reduction}, we 
  provide the individual data after applying all of
  the post-process reduction described in this section. 
  For low surface brightness measurements, 
  the sky subtraction is 
  crucial. In the final datacube, we checked that of the sky subtraction residual 
  at wavelengths outside Ly$\alpha$ emission. 
 In our final datacube, the typical 1-$\sigma$ uncertainty is 
 $\approx 9.0 \times10^{-19}$ erg s$^{-1}$ cm$^{-1}$ arcsec$^{-2}$ (1\AA\ wavelength bin). 
 The root-mean-square (rms) around zero within an aperture 
 of $2"$ is $\pm 4.5 \times10^{-19}$ erg s$^{-1}$ cm$^{-1}$ arcsec$^{-2}$  
 in the rest-frame wavelength channels between 1255\AA\ -- 1275 \AA, 
 a wavelength range that does not contain obvious emission lines from QSOs.  
 These results indicate that our sky subtraction and 
 PSF subtraction are effective.

\section{Results}

After performing the set of careful reduction steps 
described in the last section, the  
 1-$\sigma$ flux density uncertainty per voxel\footnote{voxel: a three-dimensional datapoint in an integral field spectrograph datacube.} 
 is about 1-$\sigma$ of
$\approx 3.5\times10^{-19}$ erg s$^{-1}$ cm$^{-2}$  \AA$^{-1}$ around the observed wavelength 
of 4000\AA. This measurement is consistent with the 1-$\sigma$ error 
in surface brightness we reported in the previous section. In \S2.4, we 
further checked our sky-subtraction and  PSF subtraction, and we 
confirm that the flux density in the sky-subtracted, PSF-subtracted 
cubes  is consistent with zero within the rest-frame wavelengths $\lambda=1255$ -- 1275 \AA. 
 In this section, we carefully analyze the physical properties of Ly$\alpha$ nebulae around the QSO sample at $z\approx2$.

\subsection{Optimal Extraction of the Ly$\alpha$ Nebulae}
\label{sec:extraction}

We use the three-dimensional, automatic algorithm 
 CubExtractor (Cantalupo et al. 2019) to conduct 
optimal extraction of extended Ly$\alpha$ emission
in each datacube.
We first smooth the datacubes and 
variance using a Gaussian filter with $\approx 1"\times1"$ aperture. 
Then, objects are extracted and detected if they contain 
at least six connected voxels above a signal-to-noise (S/N) 
of two after smoothing both the signal and the variance \footnote{{
The reasons of choosing six voxel limit set is the following:  To compare with previous work 
(e.g., Borisova et al. 2016; Arrigoni Battaia et al. 2019), we choose 
the number of connected voxel close to the seeing value. The number of pixels that close to 
the seeing ($\approx1''$) is $2$ voxels perpendicular to slicer direction (1.3$''$), and $3$ voxels along 
the slicer direction (0.9$''$). Thus, we use 6 voxel limit set in the manuscript. Actually, we checked 
to use 2 $\times$ 2 voxel smoothing and $3\times4$ voxel smoothing, and we found that the difference of the 
surface brightness is less than 1\%. }
}. 
The three-dimensional 
segmentation masks output by CubExtractor are then used for our analysis. 

In Figure~\ref{fig:optimal_extraction}, we present the optimally extracted images of the detected objects in each KCWI 
datacube as done by \citet{borisova16}. 
Each image has an angular size of 16.8$''\times 20''$ FoV. The images have been 
generated by (i) selecting all voxels in the PSF-subtracted and continuum-subtracted KCWI cubes,
(ii) applying the corresponding 3-D masks of each nebula output
by CubExtractor, and then
(iii) integrating the flux along the wavelength direction. 
These images are similar to pseudo-NB images, but the width of 
the filter is adjusted for each spaxel to include only signal above S/N $>2$.
The width of the pseudo-filter varies from one channel 
(typically at the edges of the object) to a few tens of wavelength channels 
in the brightest parts of the sources. The white contour in Figure~\ref{fig:optimal_extraction} 
represents the 2-$\sigma$ uncertainty of {SB $\approx1.8\times10^{-18}$ erg s$^{-1}$ cm$^{-2}$ arcsec$^{-2}$}. 
The SB uncertainty is calculated in 1 arcsec$^2$ area and in a wavelength bin of 1\AA\ (also see \citet{arrigoni19}). 
Note the 2-$\sigma$ SB limit is slightly different field-by-field. 
We summarize the SB limit for each source 
in Table~1. 

\subsection{Detection rate of giant Ly$\alpha$ nebulae}  

Above our KCWI detection limit of 
$\approx 1.8\times 10^{-18}$ erg s$^{-1}$ cm$^{-1}$ arcsec$^{-2}$, 
we find that {14 out of 16 QSOs} in our sample 
have a detected nebula with diameter $\gtrsim50$ kpc.  
Among them, Q1227, Q1228, Q1230, and Q1416 are 
enormous Ly$\alpha$ nebulae at $z\approx2$ with 
the SB$_{\rm{Ly\alpha}}>10^{-17}$ erg s$^{-1}$ cm$^{-2}$ arcsec$^{-2}$ 
for $\ge100$ kpc, 
and their projected sizes exceed the FoV of KCWI. 
Q0048 and Q1426 fields contain compact Ly$\alpha$ emitters in the 
KCWI fields, possibly indicating a strong overdense nature in these fields.

\subsection{Surface brightness of the Ly$\alpha$ Emission}

In this section, we measure and
present the radial Ly$\alpha$ profiles of the nebulae. 
For comparison with previous works, 
we use circular-averaged 
surface brightness (SB) profiles centered on the QSO continua. 
The SB profile is calculated using the pseudo-narrowband images. 
we integrate over a fixed velocity range 
of $\pm 1000$ km s$^{-1}$ around the centroid of Ly$\alpha$ 
nebular emission to calculate the surface brightness. This allows 
us to properly compare the SB calculated in our sample with 
previous work \citep[e.g.,][]{borisova16, arrigoni19}.

\begin{figure*}[h!]
\includegraphics[width=0.9\textwidth,height=0.6\textheight]{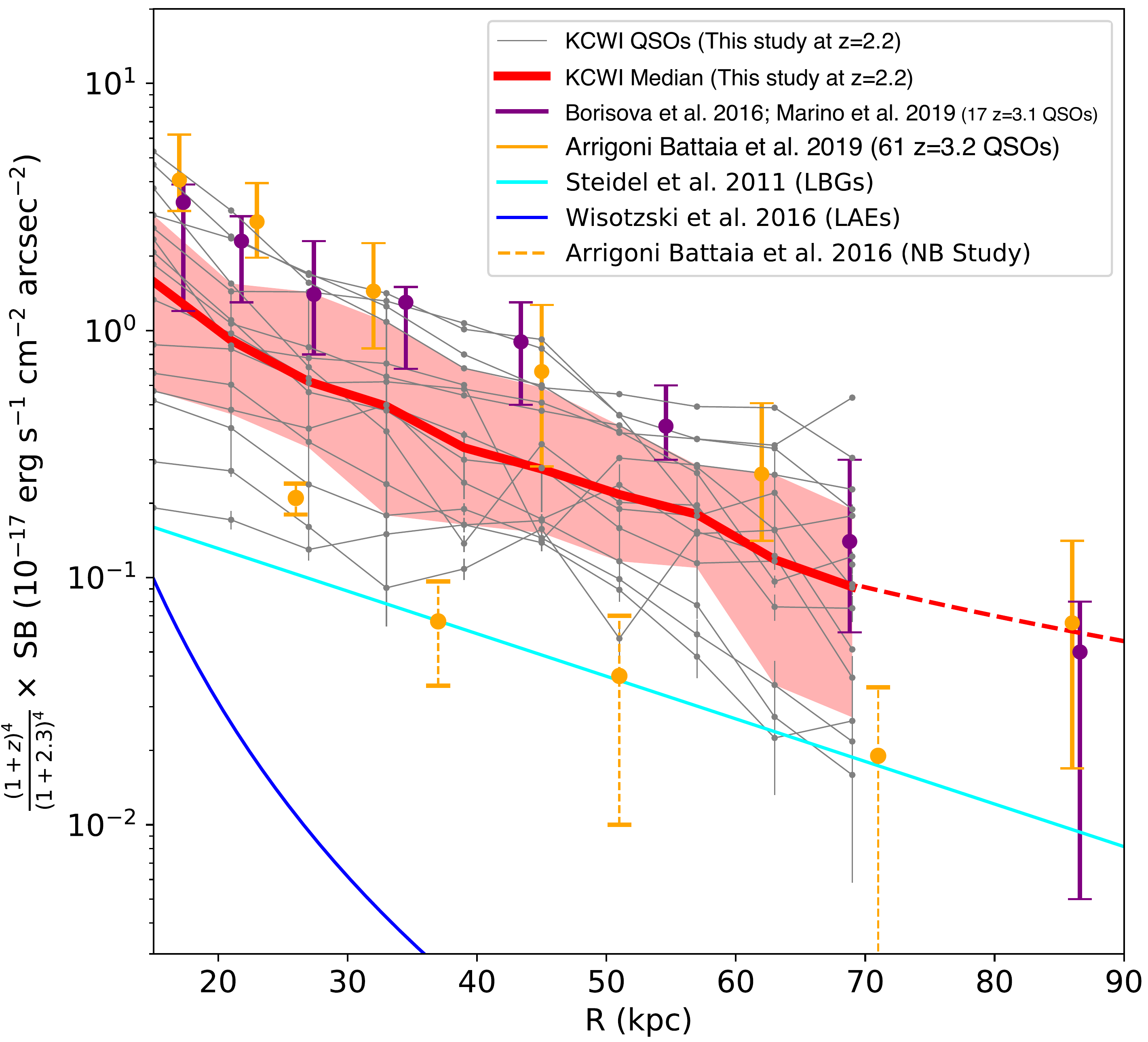}
 \caption{Ly$\alpha$ surface brightness (SB) profiles (circularly averaged) as a function of 
 radius around ultraluminous QSOs. All errorbars 
 represent the 25 -- 75 percentile 
 of Ly$\alpha$ SB in each QSO sample. The gray line with 
 cyan points represent the SB profile 
 for individual QSO at $z\approx2$. Same as \citet{borisova16} and \citet{arrigoni19}, 
 the SB is calculated in the pseudo-narrowband images with the width of 2000 km s$^{-1}$, centered on 
 the nebular Ly$\alpha$ emission (\S3.3). 
 The thick red 
 curve represents the median Ly$\alpha$ 
 profile of the KCWI sample at $z\approx2$. The solid line 
 is the median of the actual data, and the dashed red 
 curve indicates the extrapolation results. 
 The red region represents the SB within the 
 25\% and 75\% percentile of our KCWI QSO sample. 
 Purple represents the median Ly$\alpha$ SB 
 profile of 17 QSOs 
 at $z\approx3$ (Borisova et al. 2016) 
 {while the error bars represent} the 25 and 75 of 
 percentile of this sample. 
 The solid orange with error bar represent 
 the median and 25 -- 75 percentile Ly$\alpha$ SB of 61 QSOs 
 at $z\approx3.2$ reported by Arrigoni Battaia et al. (2019). 
 Orange points with dashed error bar 
 represent the SB of Ly$\alpha$ emission using 
 a narrowband filter (Arrigoni Battaia et al. 2016). 
 Cyan represents the Ly$\alpha$ 
 SB for Lyman break galaxies (LBGs) 
 at $z\approx3$ (Steidel et al. 2011), 
 and blue shows the Ly$\alpha$ profile 
 for $z\approx3$ Lyman alpha emitters (LAE)
 (e.g., Wisotzki et al. 2016). } 
 \label{fig:SB_r75}
\end{figure*}

\begin{figure*}[h]
\includegraphics[width=0.9\textwidth,height=0.6\textheight]{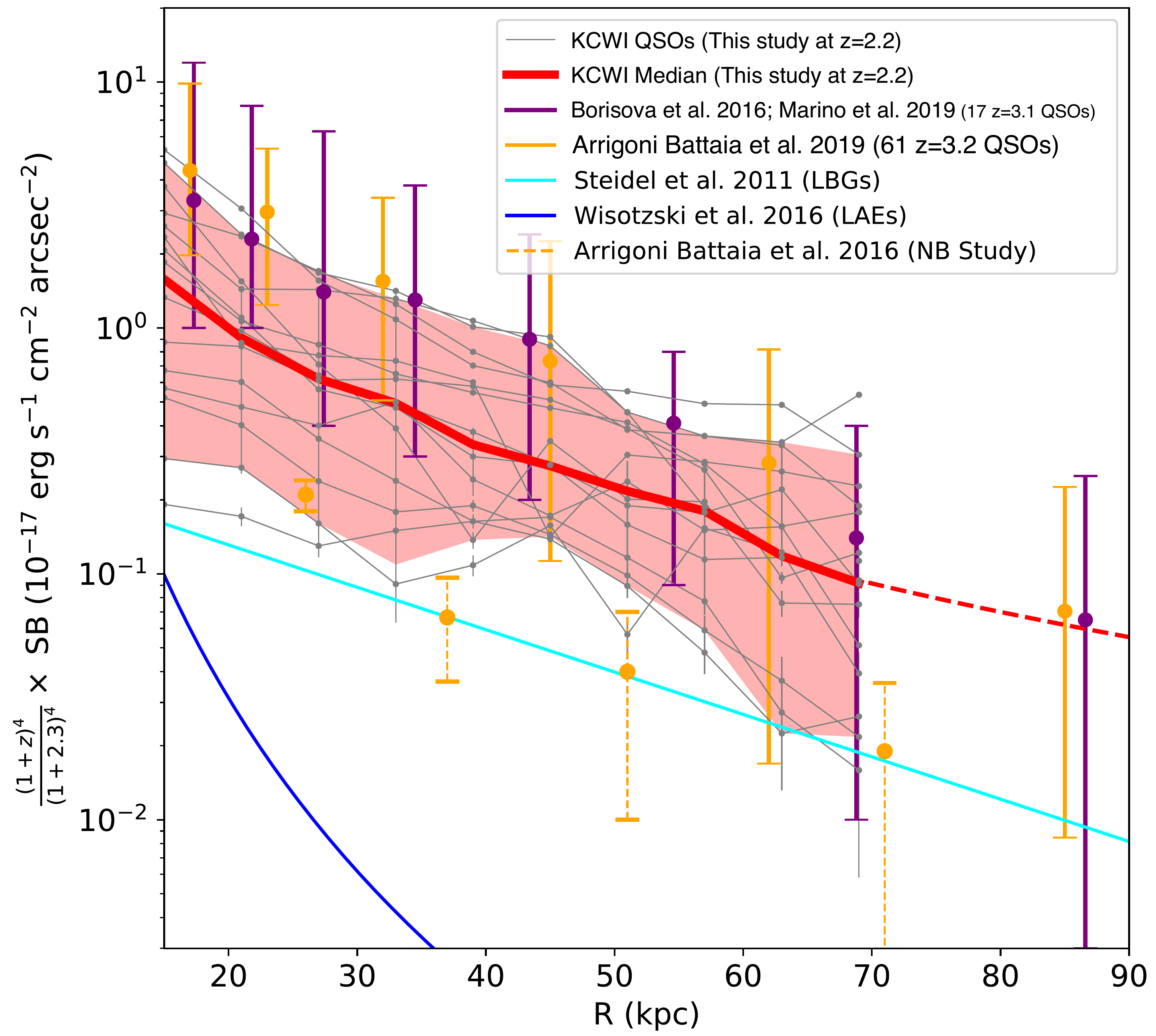}
 \caption{Similar to Fig.~\ref{fig:SB_r75}, but showing 10 -- 90 percentile of 
 the Ly$\alpha$ surface brightness of each QSO sample.} 
 \label{fig:SB_r90}
\end{figure*}

\begin{figure*}[h!]
\includegraphics[width=0.98\textwidth,height=0.7\textheight]{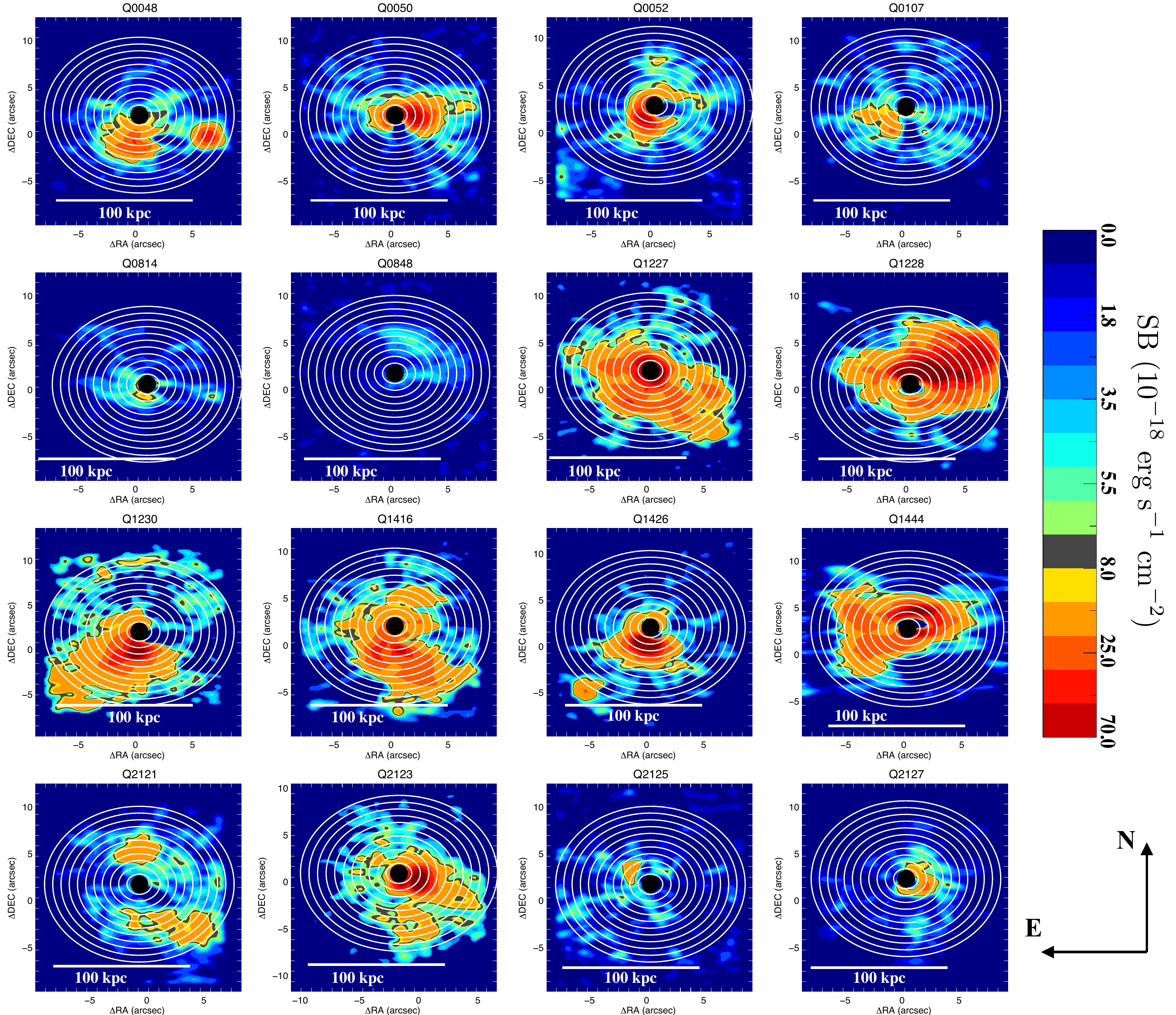}
 \caption{In this figure, we present the annuli (white circles) 
which are used to calculate the surface brightness in Figure~\ref{fig:SB_r75} 
and Figure~\ref{fig:SB_r90} for each QSO. Also, the black 
filled circle in the center represents the QSO position which is { used to calculate the circularly 
averaged SB profiles.}  } 
 \label{fig:annuli}
\end{figure*}

The individual,
circularly-averaged SB profiles for each QSOs are shown
as light gray lines in Figure~\ref{fig:SB_r75} and Figure~\ref{fig:SB_r90}. 
Further, we calculated the median SB using 
the full KCWI QSO sample at $z\approx2.2$ (shown in the thick red color 
in the Figure~\ref{fig:SB_r75} and Figure~\ref{fig:SB_r90}). 
{ In Figure~\ref{fig:annuli}, we show the annulus (white circles) 
which are used to calculate the surface brightness in Figure~\ref{fig:SB_r75} 
and Figure~\ref{fig:SB_r90} for each QSO. 
We marked the centers of each QSO position using the filled 
black circle. }
We found that the median Ly$\alpha$ SB can be described
by the following power-law profile centered at the QSO and 
valid on the radius of $r\approx15-70$ kpc:

 \begin{equation}
 \begin{split}
 SB_{\rm{KCWI}}(z\approx2.3) = & 3.7 \times 10^{-17} \times (r/40\ \rm{kpc})^{-1.8} \\
 &  \rm{erg} \ \rm{s^{-1}}   \ \rm{cm^{-2}}  \ \rm{arcsec^{-2}} 
  \end{split}
 \end{equation}

 Borisova et al. (2016) have conducted a MUSE snapshot survey 
 on 17 QSOs at $z\ge 3.1$, 
 and found that all of them are associated with 
 large Ly$\alpha$ nebula on a spatial extent of $\ge 100$ kpc.  
 Here, we denote the SB of Borisova et al. (2016) as SB$_{\rm{B}}$. 
 The median SB$_{\rm{B}}({\rm{Ly\alpha}})$ 
 can be described by the following equation: 
 \begin{equation}
 \begin{split}
 SB_{\rm{B}}(z\approx3.1) = & 3.2 \times 10^{-17} \times (r/40\ \rm{kpc})^{-1.8} \\
 &  \rm{erg} \ \rm{s^{-1}}   \ \rm{cm^{-2}}  \ \rm{arcsec^{-2}} 
  \end{split}
 \end{equation}
 

Using a larger sample of 61 QSOs at $z\gtrsim3.2$, \citet{arrigoni19} also found that 
the stacked QSO profiles can be fitted with an exponential profile with 
$SB(r) = C_{\rm{e}} {\rm{exp}}(-r/r_{\rm{h}})$ 
or a power law $SB(r) = C_{\rm{p}} r^\alpha$. 
These QSOs are characterized by a median redshift of  $z= 3.17$ ($3.03 < z < 3.46$), absolute
$i$ magnitude in the range $-29.67 \le M_i \le  -27.03$ similar 
as that of Borisova et al. (2016). 
Here, we denote $SB_{\rm{A}}$ as 
the median surface brightness derived from \citet{arrigoni19} sample. 
 \citet{arrigoni19} found that the 
obtained stacked profiles may be better fit by an exponential profile with scale length of 
$r_{\rm{h}} = 15.2\pm0.5$ kpc for radio-quiet objects, i.e., 
\begin{equation}
 \begin{split}
SB_{\rm{A}}(z\approx3.1)= & 5.4\times10^{-17} \ \rm{exp(-r/15.2 {\rm{kpc}}}) \\
&  \rm{erg} \ \rm{s^{-1}}   \ \rm{cm^{-2}}  \ \rm{arcsec^{-2}} 
  \end{split}
\end{equation}

Here, if we correct for cosmological SB dimming of MUSE \citep{borisova16, arrigoni19} 
from $z\approx 3.1$ to $z=2.2$, then the redshift-scaled median SB of the MUSE observations are:  

 \begin{equation}
 \begin{split}
 SB^{\rm{scaled}}_{\rm{B}} (z\approx2.3)\approx & 7.6 \times10^{-17} \times (r/40\ \rm{kpc})^{-1.8} \\ 
  &  \ \rm{erg}   \ \rm{s^{-1}}   \ \rm{cm^{-2}}  \ \rm{arcsec^{-2}} 
\end{split}
 \end{equation}
 
\begin{equation}
 \begin{split}
SB^{\rm{scaled}}_{\rm{A}}(z\approx2.3)\approx & 13.7\times10^{-17} \ \rm{exp(-r/15.2 {\rm{kpc}}}) \\
&  \rm{erg} \ \rm{s^{-1}}   \ \rm{cm^{-2}}  \ \rm{arcsec^{-2}} 
  \end{split}
\end{equation}
where the $SB^{\rm{scaled}}$ means that we corrected the surface brightness 
profile of both \citet{borisova16} ($SB_{\rm{B}}$) 
and \citet{arrigoni19} ($SB_{\rm{A}}$) from $z\approx3.1$ 
to $z\approx2.3$, scaled by the cosmological surface brightness dimming of the $(1+z)^4$ factor. 
{ We present the scaled data as the purple and orange points }
in Figure~\ref{fig:SB_r75} and Figure~\ref{fig:SB_r90}. 
Note the error bar is not just the statistical error, and the error bar indicates the 25\% -- 75\% percentile 
 in 
the Figure~\ref{fig:SB_r75}; and 10\% -- 90\% percentile in the Figure~\ref{fig:SB_r90} 
around the median surface brightness of $z\approx3$ QSO samples. 
The median SB of \citet{arrigoni19} is consistent with \citet{borisova16}. 
Note the median re-scaled SB of \citet{arrigoni19} is 
a factor of $1.5\times$ that of \citet{borisova16} over $20 - 40$ 
kpc in radius, and one can also see this using Eqs.(4) and (5). 
Both MUSE samples suggest that the re-scaled median SB profiles at $z\approx3$ is 
a factor of $2-3\times$ our KCWI results at $z\approx2$.

In Figure~\ref{fig:SB_r75} and Figure~\ref{fig:SB_r90}, we also include Arrigoni 
Battaia et al. (2016) results, { shown as the dashed 
orange curve with error bars}. Using narrowband observations on 
QSOs at $z\approx2$, \citet{arrigoni16} obtained the median Ly$\alpha$
SB at $z=2.2$, similar to the redshift of KCWI QSO sample. The QSOs used 
in \citet{arrigoni16} are 1.15 magnitude fainter in $i$-magnitude 
than our KCWI QSO sample. Interestingly, the narrowband Ly$\alpha$ SB is one 
order of magnitude lower than the KCWI observations. 
Note the narrow-band sample \citet{arrigoni16} indeed detected extended Ly$\alpha$ 
in 47\% of the fields with maximum projected sizes of $\lesssim50$ kpc 
above $10^{-17}$ erg s$^{-1}$ arcsec$^{-2}$. These fields are consistent with 
seven QSOs in our KCWI sample. Nevertheless, our KCWI {survey} detected a larger 
fraction of bright nebulae compared to the narrowband 
probes, yielding a much higher median SB profile. 
In \S4, we will discuss several possible reasons for this discrepancy. 
In Figure~\ref{fig:SB_r75} and Figure~\ref{fig:SB_r90}, we also plot the Ly$\alpha$ 
profile of Lyman break galaxies (LBGs) 
at $z\approx3$ \cite[e.g.,][]{steidel11} and Ly$\alpha$ emitters (LAEs) \cite[e.g.,][]{wisotzki16}. 
From the comparison, the Ly$\alpha$ profile powered by galaxies are 
much fainter than that powered by QSOs. 

\begin{figure*}[h!]
\includegraphics[width=1.0\textwidth,height=0.74\textheight]{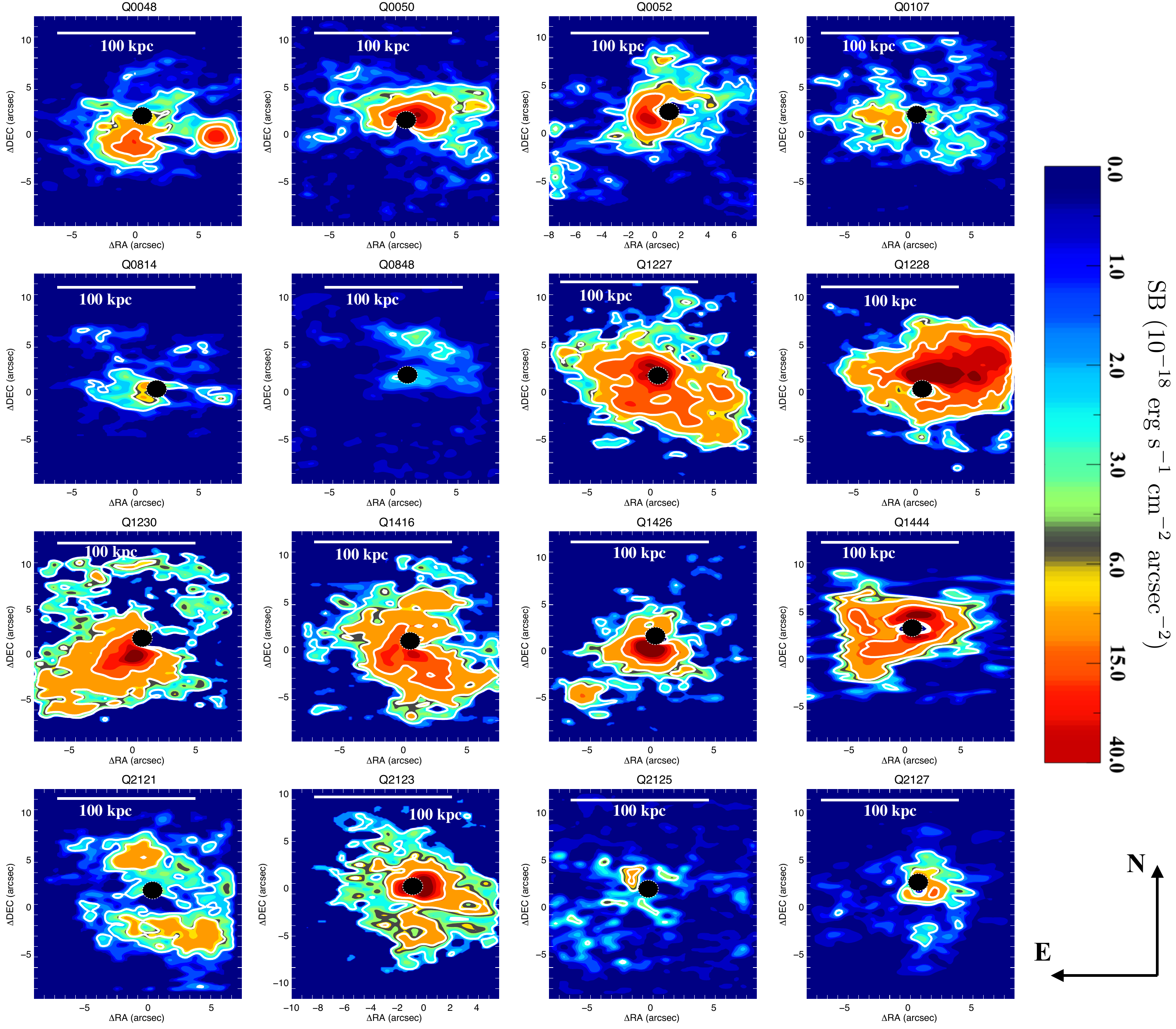}
 \caption{``Optimally-extracted" Ly$\alpha$ images from PSF and continuum-subtracted KCWI datacubes for each QSO. 
 The white bar indicates a physical scale of 
 100 kpc. Each image has a size of $16''$ in x-axis and $20''$ in y-axis. 
  The images have been produced by collapsing the datacube voxels associated with 
  the CubExtractor
 3-D segmentation maps (the ``3D-mask") along the wavelength direction (see \S2). 
 The SB is calculated using the CubExtractor 3D segmentation mask 
 that includes a different number of spectral resolution bins. 
 The 3D segmentation masks
 have been obtained with a signal-to-noise ratio (SNR) threshold of 2 per smoothed voxel, 
 similar to \citet{borisova16} and \citet{arrigoni19}. 
 {The white thick contours in each image 
 corresponds to 2-$\sigma$, 5-$\sigma$ and 10-$\sigma$ SB, with 1-$\sigma_{\rm{SB}} = 5\times10^{-19}$ erg s$^{-1}$ 
 cm$^{-2}$ arcsec$^{-2}$}. The black dots represent the positions of the central QSOs.}
  \label{fig:optimal_extraction}
\end{figure*}

\begin{figure*}[h!]
\includegraphics[width=1.0\textwidth,height=0.95\textheight]{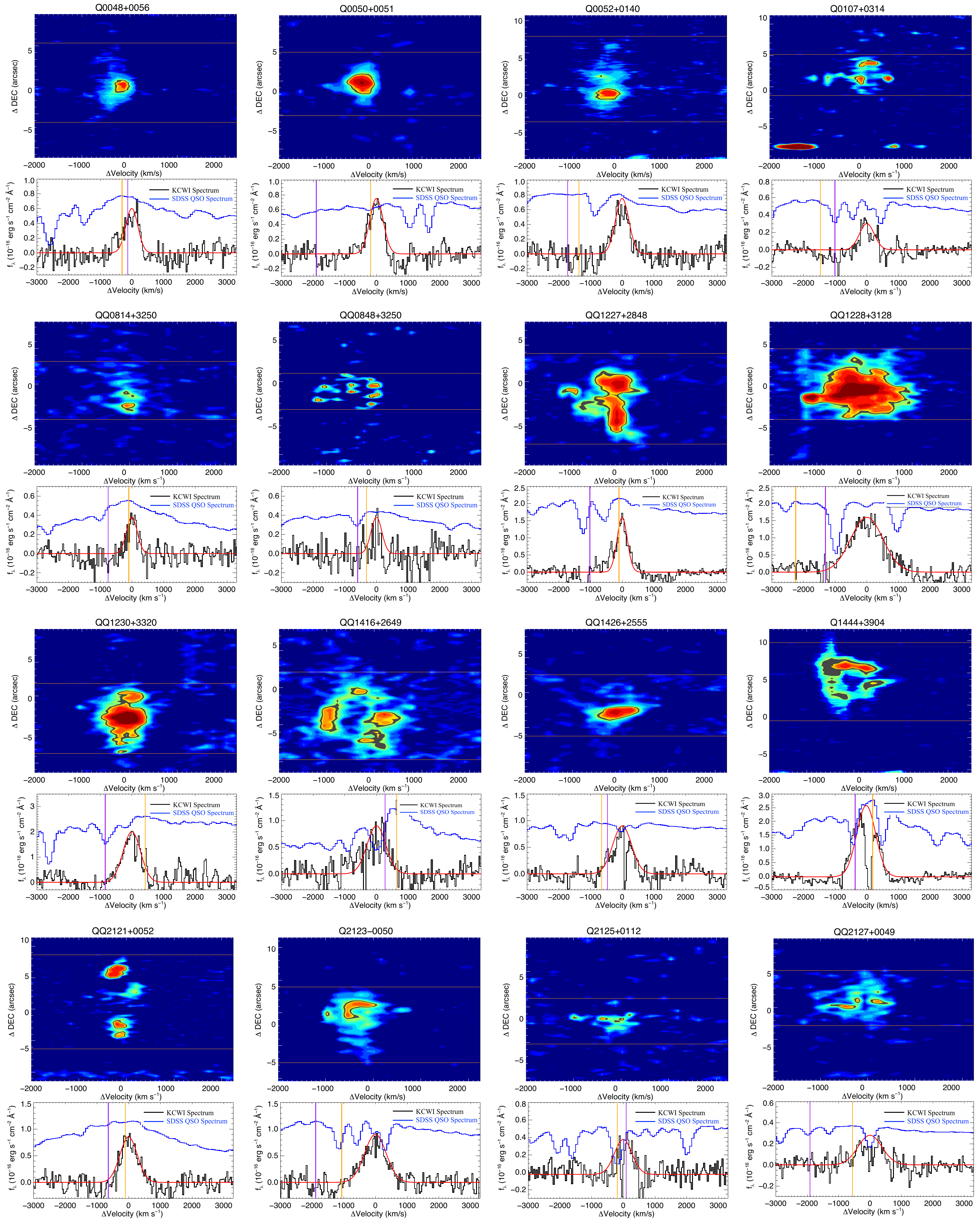}
 \caption{The 2-D and 1-D spectra of Ly$\alpha$ nebulae at $z\approx2$ 
 obtained from KCWI datacubes for each ultraluminous QSOs. 
 The upper panel shows the 2-D spectra, extracted using the pseudo-aperture defined by the thick white contours defined in Figure~\ref{fig:optimal_extraction} and integrating along the spatial x-axis direction. 
 The lower panels show the 1-D spectra of the nebulae (black lines) obtained by integrating the 1-D 
 spectra along the spatial direction between two horizontal orange lines in the upper panel.
 We overlay the one-dimensional spectrum of the QSO. Most of radio-quiet nebulae show a Ly$\alpha$ spectral shape (black) very different from that of the QSO (blue), 
 confirming that the detected emission is not an artifact of the QSO PSF subtraction. {We overplot as a vertical purple line the expected Ly$\alpha$ velocity from the systemic redshift of the QSO as determined by the SDSS pipeline fitting routine (\citealt{paris17}, also see Table~1). The orange vertical line shows instead the systemic redshift obtained by a fit of the quasar broad Ly$\alpha$ line}.} 
 \label{fig:2Dspec}
\end{figure*}

In Figure~\ref{fig:2Dspec}, we show the two-dimensional (2D) and 
the corresponding one-dimensional spectra using 
a pseudo-slit  for each QSO. The pseudo-slit for each nebula 
is the white contour shown in Figure~\ref{fig:optimal_extraction}. 
Each corresponding 1D spectrum is obtained by integrating 
all spatial pixels within the white contour in Figure~\ref{fig:optimal_extraction}. 
In each sub-figure, we detect high S/N Ly$\alpha$ emission line. 
The extended Ly$\alpha$ emission for each QSO have  
 FWHM of $400 - 800$ km s$^{-1}$, i.e.\ much narrower than 
the QSO Ly$\alpha$ emission (blue spectra in Figure~\ref{fig:2Dspec}). 
This confirms that the extended
nebular emission is unrelated to PSF
subtraction residuals. 


\begin{figure*}[h!]
\includegraphics[width=0.95\textwidth,height=0.5\textheight]{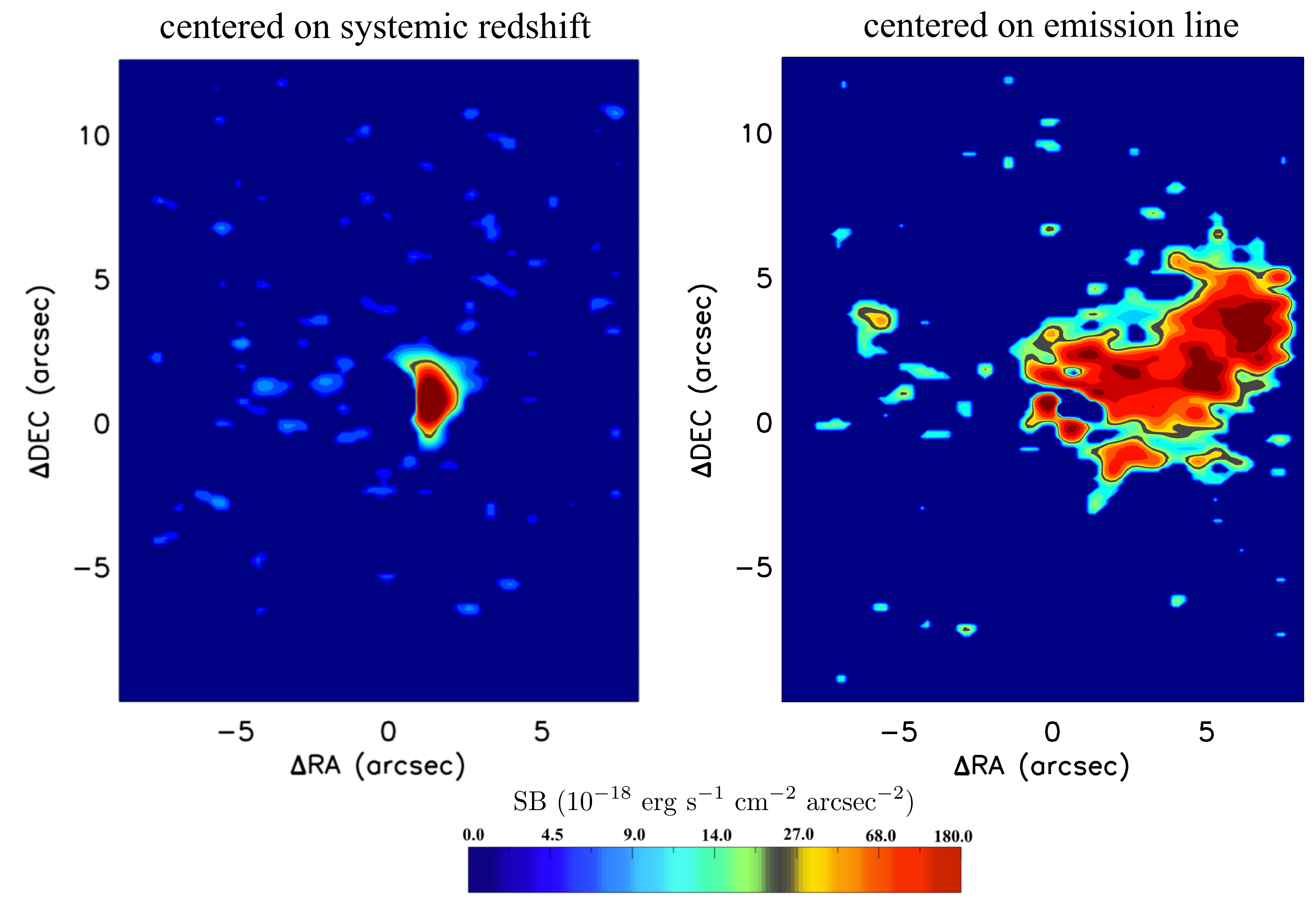}
 \caption{Comparison between two pseudo-narrowband 
  images of Q1228 {obtained with different filter central wavelengths}. 
  { The left panel shows a pseudo-narrow band image obtained assuming as the 
  filter central wavelength the systemic redshift determined using the MgII emission (Table~1)}. 
  {The right image uses instead the nebula Ly$\alpha$ redshift as the filter central wavelength.} 
  The narrowband we use has a similar profile 
  of NB3950 (e.g., Cantalupo et al. 2014), and the surface brightness is calculated by integrating 
  the flux over a wavelength bin of 30 \AA.  } 
   \label{fig:compare}
\end{figure*}

\begin{figure*}[h!]
\includegraphics[width=0.7\textwidth,height=0.44\textheight]{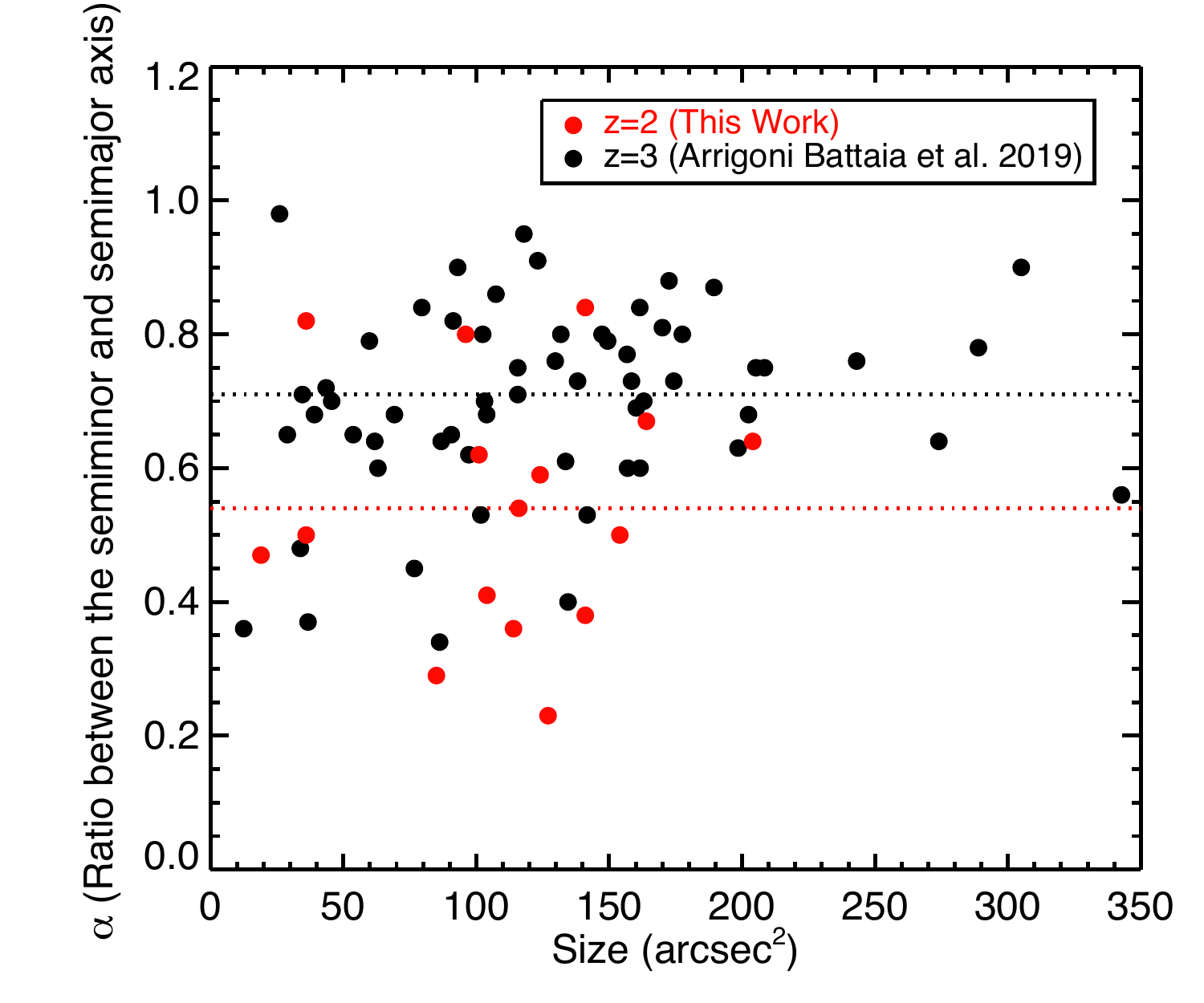}
 \caption{Plot of the asymmetry $\alpha$, 
 i.e. the ratio between the semiminor 
 axis b and semimajor axis a, 
 versus the area enclosed by the 2$\sigma$ isophote. 
 Black indicates 
 the QSOs in Arrigoni Battaia et al. (2019) at $z\approx3.1$; and the red represents QSOs in our sample at $z\approx2.2$. Our sample has a median $\alpha$ value of 0.54 (red horizontal dotted line), comparing with the median $\alpha$ of $0.71$ at $z\approx3$ (black horizontal dotted line). Although a larger sample is needed, our current data tentatively suggests that the QSOs at $z\approx2$ could be more asymmetric than that at $z\approx3$. 
 } 
 \label{fig:alpha_size}
\end{figure*}

\begin{figure*}[h!]
\includegraphics[width=1.0\textwidth,height=0.7\textheight]{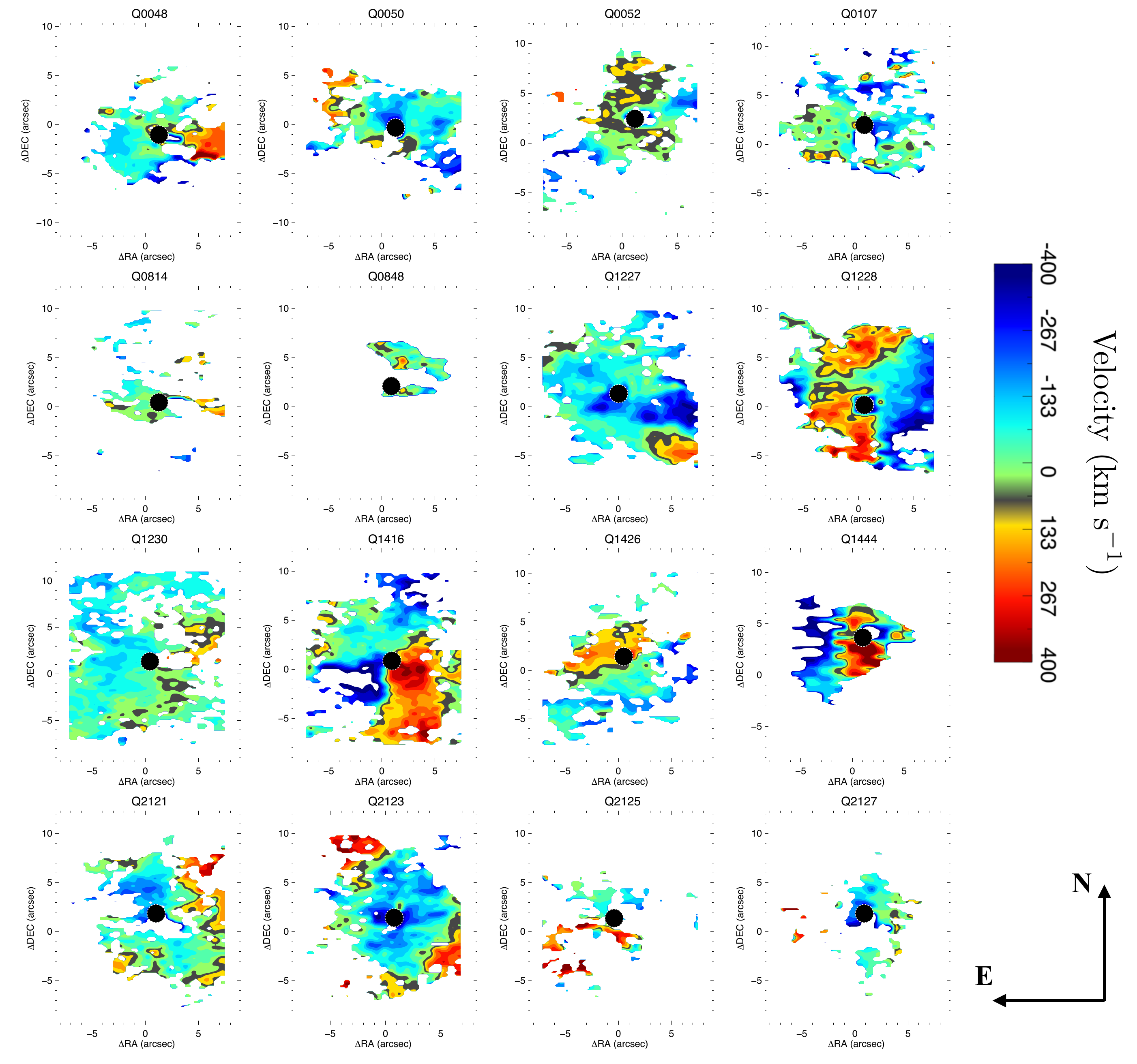}
 \caption{This figure shows the flux-weighted velocity map 
 for individual QSOs at $z\approx2$ using KCWI. } 
 \label{fig:velocity}
\end{figure*}

\begin{figure*}[h!]
\includegraphics[width=1.0\textwidth,height=0.7\textheight]{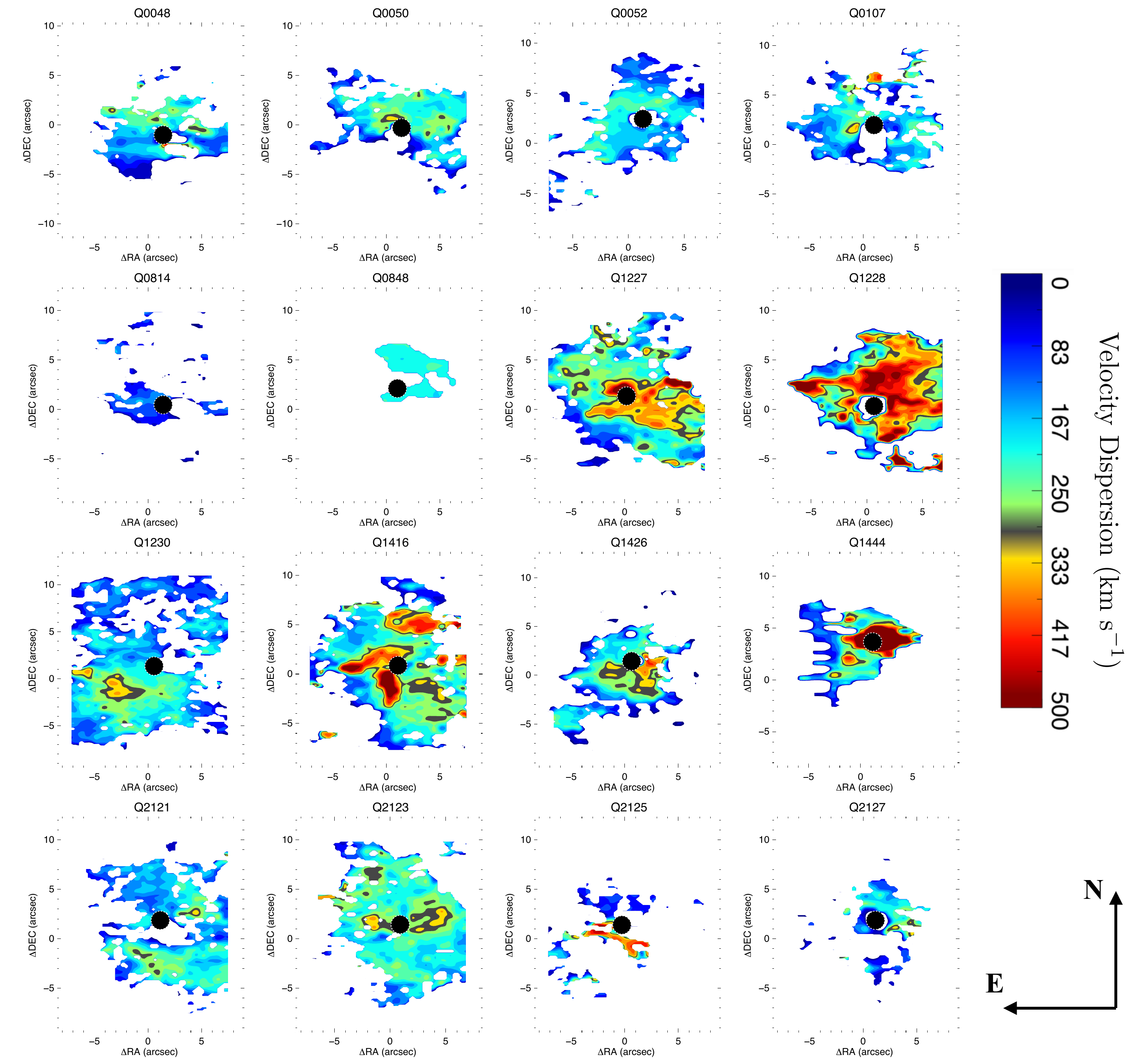}
 \caption{This figure shows the flux-weighted velocity dispersion map 
 for individual QSOs at $z\approx2$ using KCWI. } 
 \label{fig:velocityD}
\end{figure*}


\begin{figure*}[h]
\includegraphics[width=0.9\textwidth,height=0.25\textheight]{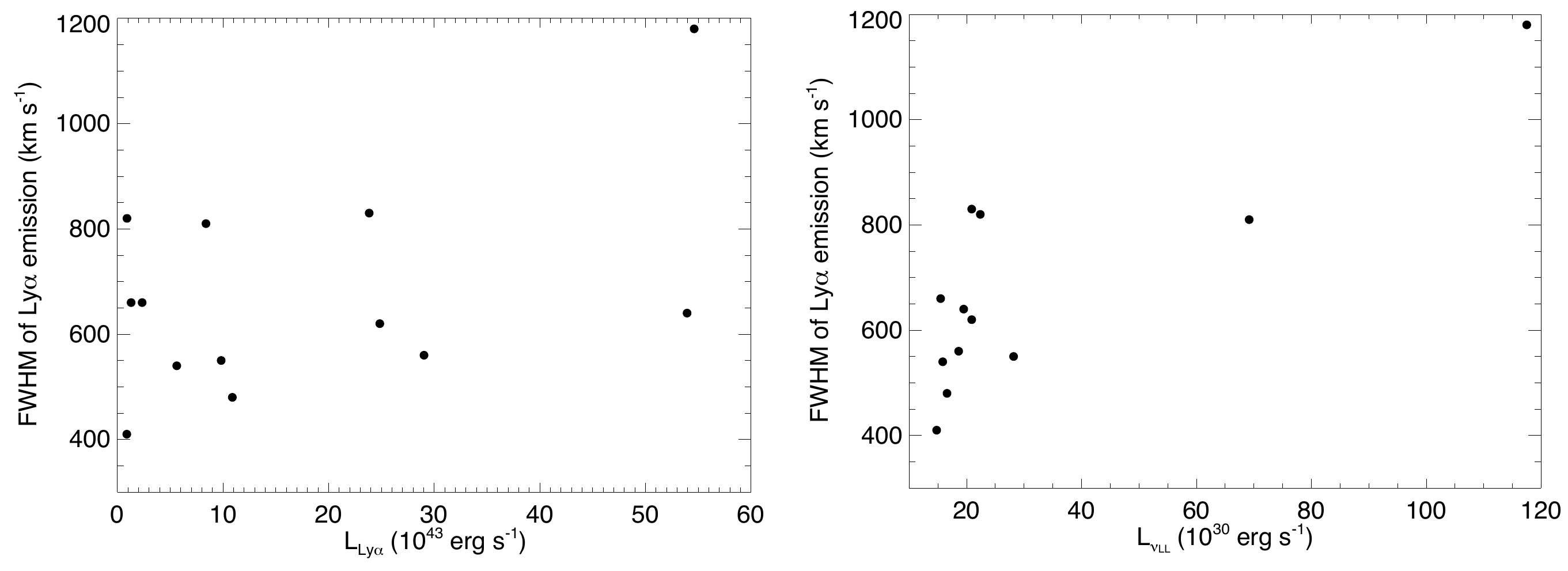}
 \caption{The left panel shows that the FWHM of Ly$\alpha$ emission as a function of nebular Ly$\alpha$ luminosity. 
 The right panel indicates the size of the FWHM 
 of Ly$\alpha$ emission as a function of ionization luminosity. Due to the extremely high 
 SNR of Ly$\alpha$ emission, the luminosity measurement has a typical error of $\le1\%$
 and the FWHM measurement has a typical error of $10\%$ (see Table~1). 
 No strong 
 correlation between the FWHM and the nebular 
 Ly$\alpha$ luminosity { is found}. } 
 \label{fig:LLya_FWHM}
\end{figure*}

\begin{figure}[h!]
\includegraphics[width=0.5\textwidth,height=0.29\textheight]{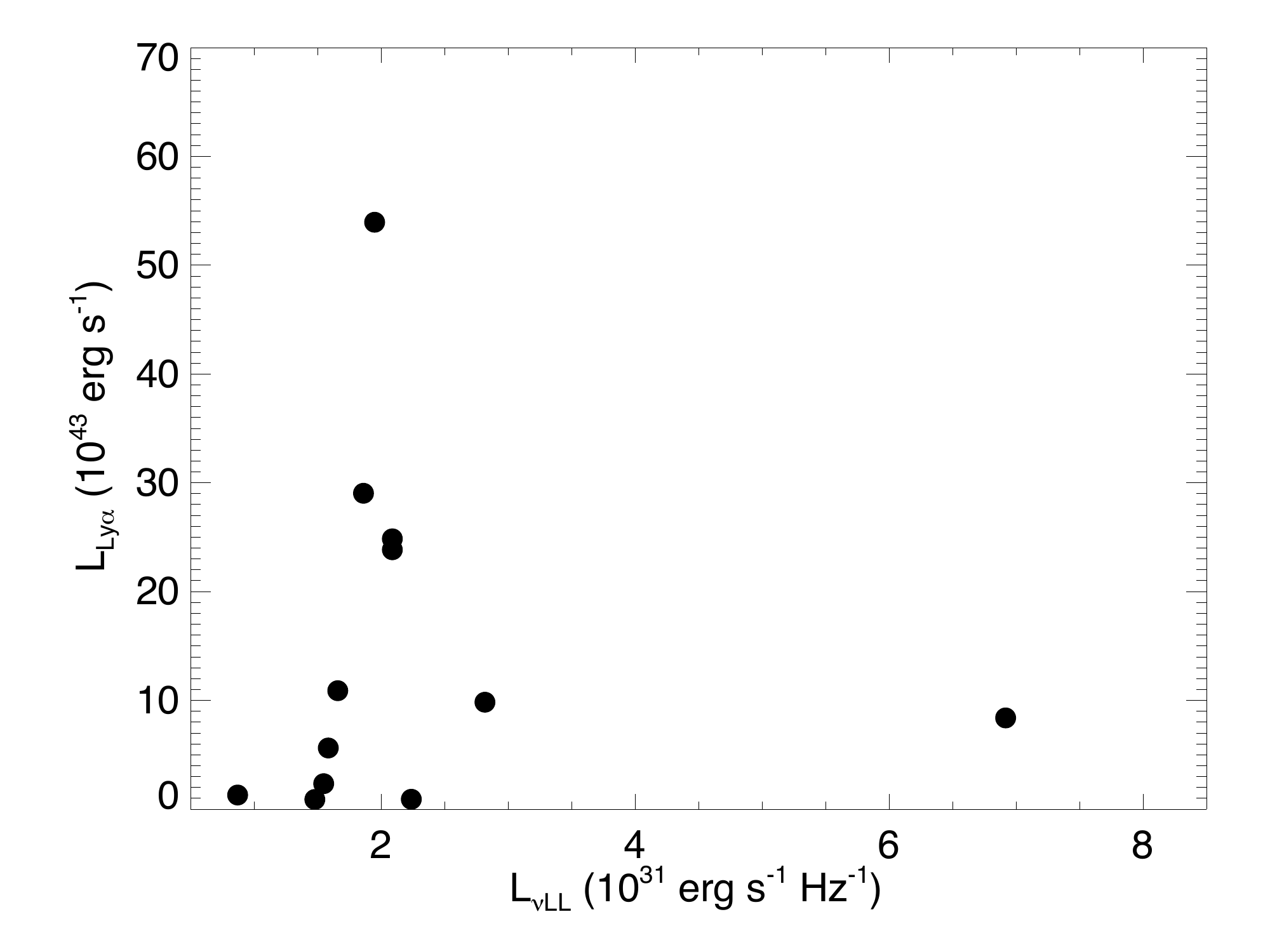}
 \caption{The relation between the Ly$\alpha$ luminosity and the QSO ionizing flux $L_{\rm{\nu}LL}$. From this figure, we can see that there is no 
 obvious correlation between $L_{\rm{Ly\alpha}}$ and  $L_{\rm{\nu}LL}$, suggesting the nebular powering 
 mechism cannot be dominated by optically thick photoionization scenario. The typical statistical 
 error of the data points in this figure is less than 1\% (see Table~1).}
 \label{fig:LLya_Lnu}
\end{figure}


\subsection{Morphology of the nebulae}
\label{sec:morphology}

From Figure~\ref{fig:optimal_extraction}, each nebula has a different morphology and size. All of them have asymmetric morphology. 
We use the method of \citet{arrigoni19}
to quantify the morphology. 
Our results can be used to quantify the morphological  evolution of the Ly$\alpha$ nebulae from $z=3$ to $z=2$. 
We used the asymmetry of $\alpha$ to 
quantify the morphology, and $\alpha$ can be calculated from the following formulae: 

\begin{equation}
\begin{split}
M_{\rm{xx}}\equiv \left\langle\frac{(x-x_{\rm{Neb}})^2}{r^2} \right\rangle_f; \ \  
M_{\rm{yy}}\equiv \left\langle\frac{(y-y_{\rm{Neb}})^2}{r^2}\right\rangle_f;  \\
M_{\rm{xy}}\equiv \left\langle\frac{(x-x_{\rm{Neb}})(y-y_{\rm{Neb}})}{r^2}\right\rangle_f
\end{split}
\end{equation}
where $x_{\rm{Neb}}$, $y_{\rm{Neb}}$ 
are the flux-weighted centroid 
for each nebulae within the 2-$\sigma$ isophoto in the 2D 
image shown in Figure~\ref{fig:optimal_extraction}, and $r$ is the distance of a point (x,y) from the flux-weighted centroid. The subscript $f$ 
represents the flux-weighted centroid. 
\begin{equation}
Q\equiv M_{\rm{xx}} - M_{\rm{yy}}, \ \ U\equiv 2 M_{\rm{xy}}
\end{equation}
where M$_{\rm{xx}}$, M$_{\rm{xy}}$, M$_{\rm{yy}}$ are called 
second-order moments (``Stokes parameters") (see Arrigoni Battaia et al. 2019). 
Using Eq.~(6) and Eq.~(7), we define the asymmetry $\alpha$ using the following equation: 

\begin{equation}
\alpha = b/a = \frac{(1- \sqrt{Q^2+U^2})}{1+ \sqrt{Q^2+U^2}}
\end{equation}
We find that at $z \approx 2 $, the median of the asymmetry 
of our 16 QSOs is 0.54, with a 1$\sigma$ scatter of 0.18 (see Figure~\ref{fig:alpha_size}). Note the 
surface brightness for yielding this results are 2$\sigma$ limit of $\approx 2\times10^{-18}$ erg s$^{-1}$ cm$^{-2}$ 
arcsec$^{-2}$. 
The asymmetry of the nebular morphology does not have a strong dependence of the size of the nebulae. 
The bright nebulae with 
diameters $>100$ kpc (e.g., Q2123, Q1230, Q1416) show clumpy and filamentary structures. 
Q2121 shows two major components. Q0048 shows a strongly asymmetric Ly$\alpha$ distribution and 
a compact Ly$\alpha$ emitter in the field. 
For a comparison, \citet{borisova16}
suggest that most nebulae 
at $z>3$ with modest Ly$\alpha$ emission have more 
symmetric and circular emission; nebulae with scales 
$>200$ kpc show evidence of filamentary structure 
and complex multiple components. \citet{arrigoni19} also found that 
most of the nebulae at $z>3$ have symmetric and round morphologies, 
with a median asymmetry of $\alpha=0.71$. One of the most asymmetric 
structure in the sample is the ELAN in \citet{arrigoni19} with $\alpha\approx0.5$. 
The $z\approx2$ QSOs have a smaller median asymmetry. Although larger sample is required, 
$z\approx2$ QSOs may be more asymmetric and clumpy  
compared to the $z\approx3$ samples \citep{arrigoni19}. We will discuss 
the implications of these observations in \S4.5.  

\subsection{Kinematics of the Ly$\alpha$ Nebula}

Although the Ly$\alpha$ line may be broadened by radiative transfer effects, 
 the relative comparison between the kinematics of 
different objects is still informative. 
In Figure~\ref{fig:velocity}, we show the map of the first moment (flux-weighted velocity) 
for each Ly$\alpha$ nebula, centered on 
the peak of the integrated Ly$\alpha$ emission of each nebula. 
Q1416, Q1228, may show possible rotation kinematics in 
 disk-like structures, possibly 
suggesting kinematics of gas inflow predicted by simulations
\cite[e.g.,][]{stewart13}. 
Nevertheless, the majority of the nebulae do not show any clear evidences 
 of rotation or well-regulated kinematic patterns. 


In Figure~\ref{fig:velocityD}, we present the velocity dispersion map 
(the second moment of the flux
distribution). 
The median value of the velocity dispersion 
for each sources ranges from 83 -- 381 km s$^{-1}$ \footnote{Note the FWHM of 
the emission line if Gaussian  
is approximately a factor of 2.35 $\times$ 
the velocity dispersion. }.
Our current data does not suggest  
any clear correlation between the luminosity and 
the velocity dispersion of Ly$\alpha$ emission (Figure~\ref{fig:LLya_FWHM}). 
The velocity dispersion shown in Figure~\ref{fig:velocityD} is consistent 
with the FWHM calculated from the integrated spectra shown in the Figure~\ref{fig:2Dspec}. 
The velocity dispersion of $z\approx2$ QSOs have a consistent range 
with that of $z\approx3$ \citep{borisova16, arrigoni19}.
The motions within Ly$\alpha$ nebulosities have amplitudes 
consistent with gravitational motions expected in dark matter halos hosting 
QSOs ($M_{\rm{halo}}\sim10^{12.5}\ M_\odot$) at these redshifts \cite[e.g.,][]{arrigoni19}. 


\section{Discussions}
\label{sec:dicussion}

From the analysis presented in the \S3, it is
clear that our KCWI observations 
reveal extended Ly$\alpha$ emission on the projected 
scales exceeding 50 kpc around 14 out of 16 of 
the $z\approx2$ QSOs (except Q0848 and Q2125). 
Further, the median surface brightness 
of Ly$\alpha$ in the KCWI study is higher than that of  
previous narrowband  studies at similar redshift. 
 KCWI results suggest that the typical 
circularly-averaged SB profile of Ly$\alpha$ nebulae
at $z\approx2$ is a factor of $2-3$ lower than that at $z\approx3$ determined 
by MUSE observations \cite[e.g.,][]{borisova16,arrigoni19}. 
In this section, we discuss the implications of the observations, 
relying only on the information
coming from the KCWI observations.

 \subsection{A much higher detection rate of Ly$\alpha$ nebulae at $z\approx2$ 
 compared to narrowband surveys}
 
The results of the KCWI survey at $z\approx2$ is in contrast with previous 
narrowband studies at the same redshift. 
In our KCWI survey, 
{ 14 out of 16 QSOs} at $z\approx2$ are 
associated with Ly$\alpha$ nebulae having 
projected linear scales of $>50$ kpc, 
with the 2-$\sigma$ SB limit of 
$2\times10^{-18}$ erg s$^{-1}$ cm$^{-2}$ arcsec$^{-2}$. 
The median profile is 
$SB_{\rm{Ly\alpha}}= 3.7 \times 10^{-17} \times (r/40\ \rm{kpc})^{-1.8}$ 
erg s$^{-1}$ cm$^{-2}$ arcsec$^{-2}$ (see \S3).  

The KCWI survey reaches a typical 2-$\sigma$ SB of $1.6\times10^{-18}$ erg s$^{-1}$ 
cm$^{-2}$ arcsec$^{-2}$ (1 \AA\ wavelength bin), 
deeper than previous narrowband surveys (typically 
$\approx4\times10^{-18}$ erg s$^{-1}$ cm$^{-2}$ in \citet{arrigoni16}). 
The luminous, large Ly$\alpha$ emission  
detected in our KCWI survey 
is rarely found in previous     
narrowband survey.
We use dashed yellow error bars in (Figure~\ref{fig:SB_r75} and Figure~\ref{fig:SB_r90}) 
to represent a large narrowband survey 
(Arrigoni Battaia et al. 2016) which study 
suggests that the Ly$\alpha$ SB level is $5\times10^{-19}$ erg s$^{-1}$ cm$^{-2}$ arcsec$^{-2}$ 
at $R\approx50$ kpc. The median SB probed by KCWI  
is about one order of magnitude brighter than that of previous 
narrowband survey shown in dashed yellow line in Figure~\ref{fig:SB_r75} and Figure~\ref{fig:SB_r90}. 

Let us explore several possibilities that could result in 
significantly fainter Ly$\alpha$ 
profiles that were obtained 
by narrowband 
surveys at $z\approx2$. 
We propose the following explanations: 
(1) light loss due to the narrowband filter;  
(2) fainter QSOs due to smaller QSO sample constrained by the narrowband. 

 The first reason could be due to the Ly$\alpha$ line falling 
 outside (or partially outside) the filter.
 In Figure~\ref{fig:2Dspec}, we marked the systemic redshift of these 
 QSOs using vertical  
 purple lines. 
 Similar to \citet{arrigoni19}, for our KCWI sample, we  
used the systemic redshift determined from SDSS \mgii\ emission, correcting 
the luminosity-dependent offset between \mgii\   and 
systemic redshift (Richards et al. 
2002; Shen et al. 2016). 
Our QSO sample resides at the bright-end in the SDSS QSO sample, 
and all of them have a high SNR \mgii\   emission detected in the 
SDSS spectra. 
 Note that all QSOs except Q1416, have large systemic 
 redshift offset comparing to that of the nebular Ly$\alpha$ emission. 
More than 50\% of the QSOs have systemic redshifts  
$\gtrsim 1000$ km s$^{-1}$  bluer than the Ly$\alpha$ 
redshift of the extended nebulae.
80\% of the QSOs have $\gtrsim 800$ km s$^{-1}$ velocity offset between 
nebular Ly$\alpha$ emisison and QSO systemic redshift determined by SDSS pipeline ($z_{\rm{pipe}}$ in Table~1). 
{ If we use the redshift of principle component analysis (PCA) $z_{\rm{PCA}}$ (e.g., \citealt{paris17}), we still 
found that 10 out of 16  have velocity offset $>500$ km s$^{-1}$ away from the nebular Ly$\alpha$ redshift. }
When conducting 
narrowband surveys, if we require the QSO systemic redshift 
to be within 500 km s$^{-1}$ from the most sensitive part of 
the filter, then most of nebular emission would be outside the sensitive 
part of the filter. For example, the narrowband 
filter that Arrigoni Battaia et al. (2016) used has a FWHM of 3000 km s$^{-1}$. Thus, 
a velocity offset of $>1300$ km s$^{-1}$ can significantly 
  reduce the observed Ly$\alpha$ flux and decrease the ability to 
  detect extended emission. { In the Figure~\ref{fig:compare}, we present a simulated 
  narrowband image of Q1228}. The systemic redshift of Q1228 is 600 km s$^{-1}$ 
  bluer than the 
  Ly$\alpha$ redshift. 
  Using 
  a narrowband with FWHM of $\approx 30$ \AA\ centered on the systemic 
  redshift (the profile of the filter is the same as the filter used in Arrigoni 
Battaia et al. 2016), the Ly$\alpha$ emission is very compact. 
  { From Figure~\ref{fig:compare}, if the narrowband filter has a central wavelength
consistent with the systemic redshift (e.g., Arrigoni Battaia et al. 2016),}
Ly$\alpha$ is a compact source which is easily missed under the extremely bright PSF in the narrowband imaging. 
The IFU, compared to narrowband, 
does not have the narrow redshift constraint, and thus, 
one can always detect nebular Ly$\alpha$ emission even if
the  velocity offset between 
nebular Ly$\alpha$ and QSO systemic is larger than expected.

 

Due to the wide wavelength coverage afforded by KCWI, 
the KCWI QSO sample is 1.15-mag brighter
than the narrowband QSO sample 
(Arrigoni Battaia et al. 2016).
However, using our sample, we demonstrate that there 
are no obvious correlation between nebular Ly$\alpha$
and QSO ionizing flux (or QSO magnitude) (see Figure~\ref{fig:LLya_Lnu}). Thus, 
there is no strong evidence to support that the 1.15-mag 
fainter QSO sample to be a major factor for the one order 
of magnitude discrepancy between the extended nebular Ly$\alpha$ 
observed by KCWI and narrowband (also see the similar 
conclusion for $z\sim3$ in \citet{arrigoni19}). 


\subsection{Evolution of the Circularly-Averaged Ly$\alpha$ Surface Brightness 
from $z\gtrsim3$ to $z\sim2$}

   From Eq.~(1) and Eq.~(3), Figure~\ref{fig:SB_r75} and Figure~\ref{fig:SB_r90}, 
   we know that 90\% of the $z\approx2$ Ly$\alpha$ nebulae have circularly averaged SB 
   profiles fainter than the median SB profiles 
   Ly$\alpha$ SB at $z\gtrsim3$ (Borisova et al. 2016; Arrigoni Battaia et al. 2019; also see Marino et al. 2019). 
   The median SB at $z\approx2$ is 0.4 dex fainter than the median 
   Ly$\alpha$ SB profile at $z\approx3$. 
   In this section, we investigate 
   possible cause of such an evolution. 
   
 Comparing with $z\approx3$, the lower circularly averaged SB at $z\approx2$ 
can be arise from two scenarios: 
(a) Nebulae at $z\approx2$ could have less circular morphology 
or lower emission cover fraction than that at $z\approx3$. 
(b) The nebular SB are intrinsically fainter. This could further suggest  
 lower local densities or lower gas mass in the QSO halos at $z=2$ compared to $z=3$. 
Our measurement is based on a sample of 16 QSOs at $z\approx2$, and 
a larger $z\approx2$ QSO sample should be required to further confirm 
the results in this paper.

\subsubsection{Comparison between the Covering Fraction of the Ly$\alpha$ 
Emitting Cloud at $z\gtrsim3$ and $z\approx2$}

For scenario (a), we quantify the 
redshift-corrected area of a characteristic 
surface brightness. We choose the characteristic 
surface brightness of $1\times10^{-17}$ erg s$^{-1}$ 
cm$^{-2}$ arcsec$^{-2}$ which is well detected in each of 
our KCWI QSOs at $z\approx2.3$ and 
represented in orange color in Figure~\ref{fig:optimal_extraction}. The fiducial  
SB at $z\approx2.3$ is corresponding to $\approx 
4\times10^{-18}$ erg s$^{-1}$ cm$^{-2}$ arcsec$^{-2}$ 
at $z\approx 3.1$. This value is well detected for 
each QSOs at $z\approx3.1$ in MUSEUM QSO survey 
(Arrigoni Battaia et al. 2019). For all nebulae at $z\approx3$, 
the SB contours of 
$\gtrsim4\times10^{-18}$ erg s$^{-1}$ cm$^{-2}$ arcsec$^{-2}$ 
at $z\approx3.1$ have projected size within the KCWI FoV 
(Arrigoni Battaia et al. 2019), and thus, 
the area at $z\approx2$ and $z\approx3$ can be directly 
compared without correcting the difference of the FoV between 
KCWI and MUSE. 

\begin{figure}[h]
\includegraphics[width=0.55\textwidth,height=0.32\textheight]{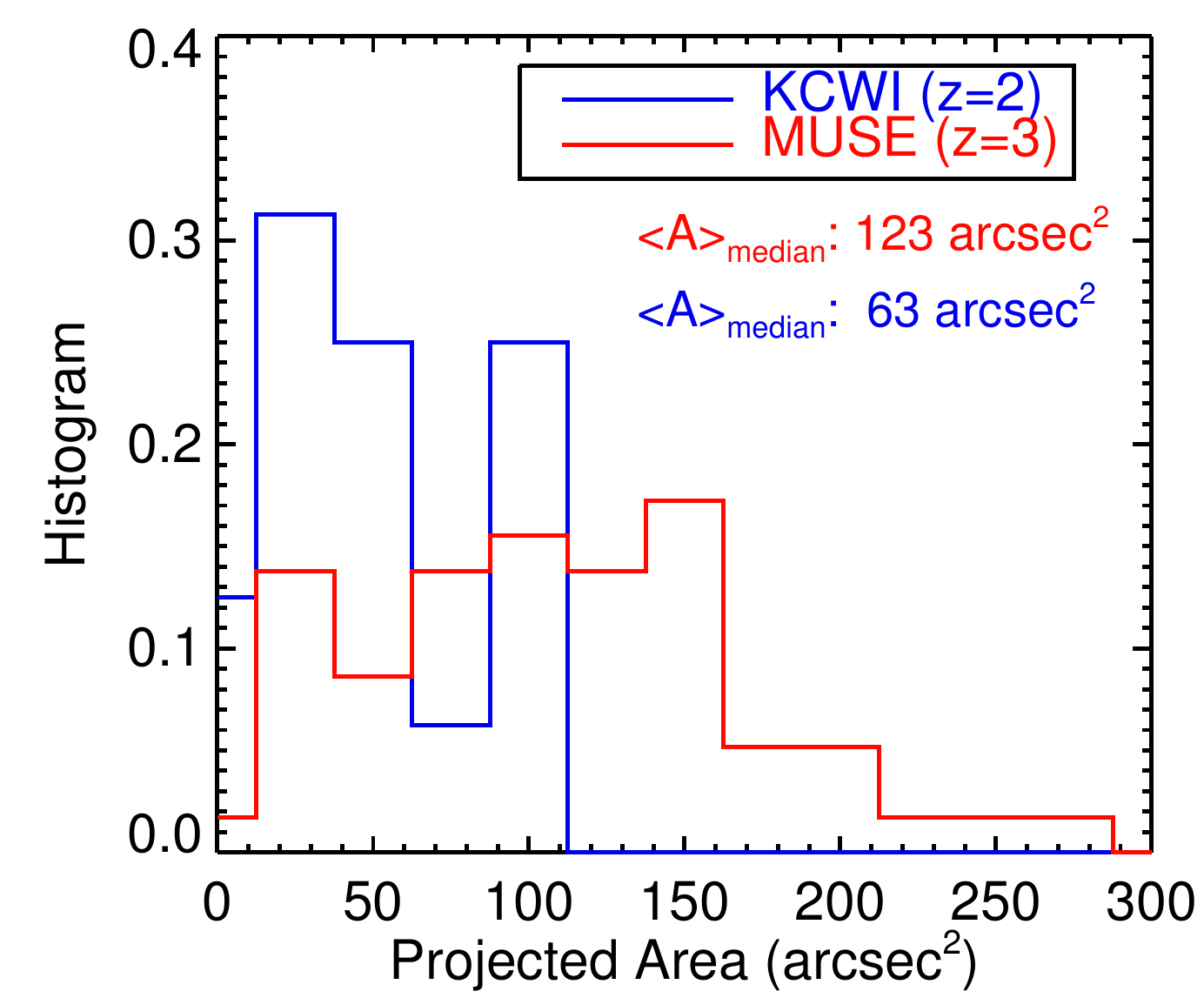}
 \caption{The distribution of the projected area with redshifted 
 surface brightness (SB$_{\rm{z=2.3}}$) greater than $1.0\times10^{-17}$ 
 erg s$^{-1}$ cm$^{-2}$ arcsec$^{-2}$. Because all MUSE nebulae at $z\approx3$ 
 have redshifted SB$_{\rm{z=2.3}}\gtrsim 1.0\times^{-17}$ erg s$^{-1}$ 
 cm$^{-2}$ arcsec$^{-2}$ contour smaller than the KCWI FoV ($16"\times20"$),
 the area at $z\approx2$ and $z\approx3$ can be directly compared without correcting the 
 difference of the FoVs between KCWI and MUSE. The median projected area at $z=2.3$ 
 is 63 arcsec$^2$ which is about 51.3\% of the median 
area of $A_{\rm{median}}=123$ arcsec$^2$) for nebulae at $z\approx3.1$. } 
 \label{fig:area}
\end{figure}

In Figure~\ref{fig:area}, we present the projected 
area of the nebulae with the SB above the fiducial value, 
and compared with $z=3$ MUSE results (Arrigoni Battaia et al. 2019) 
with the same scaled SB at $z=2$. 
From the figure, the median area at $z=2$ is 
about 63 arcsec$^2$ (4340 kpc$^2$), this is about 55.1\% of the median 
area of MUSE nebulae of $A_{\rm{median}}=123$ arcsec$^2$ (7872 kpc$^2$). 
Let us further take into account the halo size from $z\approx3$ -- $z\approx2$. 
Let us assume the QSO halo to have a halo mass of $M_{\rm{halo}}\approx10^{12.5}\ M_\odot$ 
(e.g., White et al. 2012; Shen et al. 2007) at both redshifts. 
 Under this assumption, the Virial radius of the halo is $R_{\rm{h, z=3.1}}\approx112$ kpc 
 at $z\approx 3.1$, and $R_{\rm{h, z=2.3}}\approx138$ kpc at $z\approx2.3$. 
 The {covering} factor of Ly$\alpha$ emitting region 
 can be estimated using 4340 kpc$^2$ / ($\pi R_{\rm{h, z=2.3}}$)$\approx0.07$ 
 at $z\approx2.3$. Using the same method, we estimate that the {covering} factor  
 of the Ly$\alpha$ emitting at $z\approx3.1$ is about 0.19, a factor of $2.7\times$ that at $z\approx2.3$. 
Therefore, we conclude that QSO halos at $z\approx2$ have a lower 
covering factor of the Ly$\alpha$ emitting clouds compared to that at $z\approx3$, 
and the lower covering factor may be one of the main reasons for the 
evolution of circular-averaged Ly$\alpha$ SB from $z=3$ to $z=2$. 

\begin{figure*}[h!]
\includegraphics[width=1.0\textwidth,height=0.3\textheight]{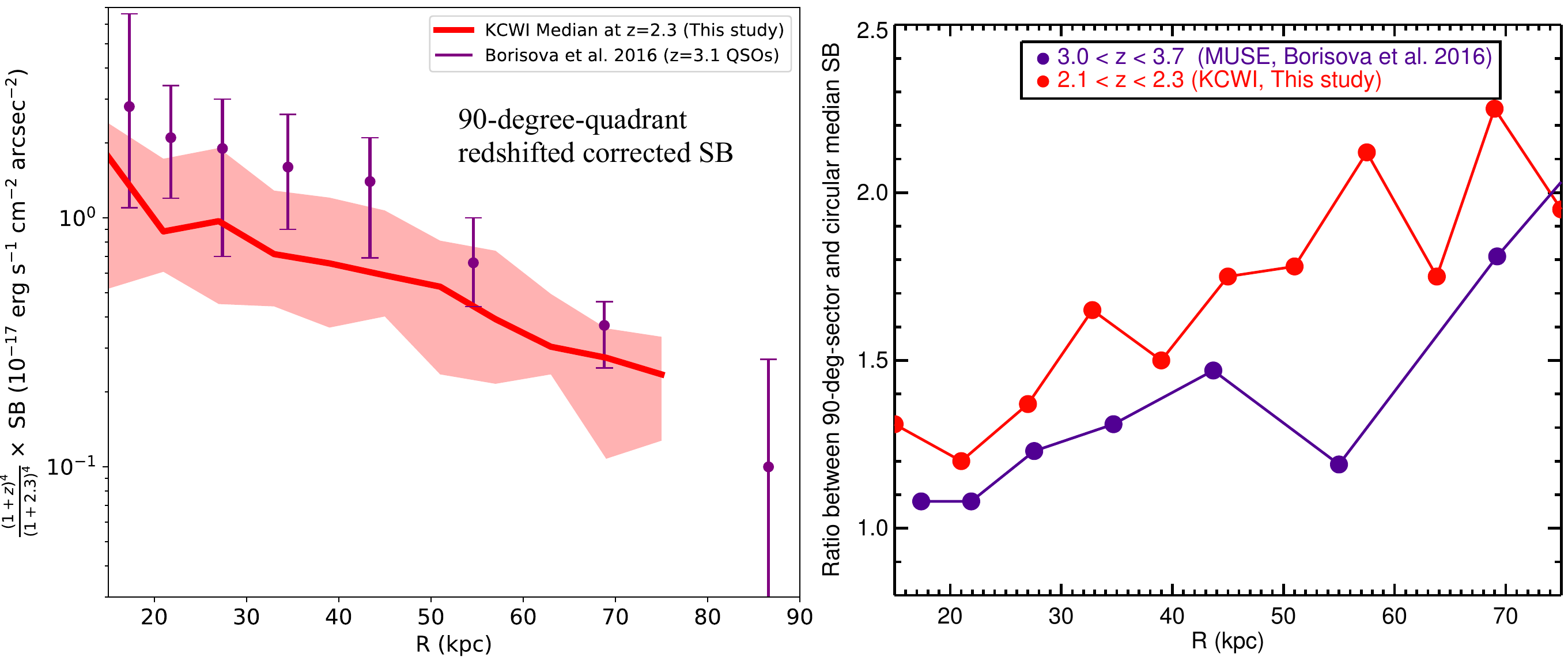}
 \caption{{\it Left:} {SB profiles averaged over a 90-degree quadrant chosen as the one with the highest flux}. {\it Right:} the ratio between the 
SB in 90-degree-sectors and the median SB in full annuli. 
On 15 -- 70 kpc, the ratio at $z\approx2.3$ (red) is 
higher than that at $z\approx3$ (purple), 
suggesting that the covering factor of the Ly$\alpha$ emitting 
clouds at $z\approx2$ is smaller than $z\approx3$ in these 
two samples.  } 
 \label{fig:SB_sector_corr}
\end{figure*}

We can further use another independent method to 
confirm that the lower circularly-averaged SB at $z\approx2$
seen in Figure~\ref{fig:SB_r75} and Figure~\ref{fig:SB_r90} is due to the lower covering factor. 
{We calculated} the median SB in a 90-degree-quadrant for each nebula. 
\footnote{Quadrant: a sector of 90 degree, and the quadrant 
is chosen with the highest median SB over a large number 
of rotations.} Then, we compute the ratio of the 90-degree-quadrant 
SB to the full annuli SB. A higher value of this ratio indicates 
a lower covering factor or a more asymmetric distribution of the Ly$\alpha$ emitting regions. 
In the right panel of Figure~\ref{fig:SB_sector_corr}, 
we plot the ratio of the 
SB in a 90-degree-sector \footnote{Again, for each QSO, the quadrant 
is chosen with the highest median SB over a large number 
of rotations.} to the median SB in the full annuli. 
We could see that this ratio at $z\approx2.3$ (red) is 
higher than that at $z\approx3$ (purple) on the scale of 15 -- 70 kpc, 
suggesting that the covering factor of the Ly$\alpha$ emitting 
clouds at $z\approx2$ is lower than that at $z\approx3$. 
In the left panel of Figure~\ref{fig:SB_sector_corr}, we 
further show the 90-degree-quadrant, redshift-corrected 
SB at $z\approx2.3$. Compared to the SB calculated using full 
annuli, the $z\approx2$ and $z\approx3$ SB are more 
consistent with each other, indicating that the characteristic, local Ly$\alpha$ SB  
within the Ly$\alpha$ emitting regions at $z\approx2$ is close to that at $z\approx3$. 
The dimming 
of the circularly average SB at $z\approx2$ is due to a smaller 
covering factor of Ly$\alpha$ emitting clouds. 

\subsubsection{Evolution of the Cosmological Gas Density}
  
   From the above section, we have shown that at the fixed SB, our KCWI QSO 
   sample at $z\approx2$ has less area covered by Ly$\alpha$
 emitting regions compared to QSOs  at $z\approx3$ \citet{borisova16, arrigoni19}.
 In the following, we investigate the scenario (b): whether the $z\approx2$ nebulae 
 are intrinsically fainter, i.e., whether $z\approx2$ QSOs have 
 lower CGM densities at $z=2$ compared to $z=3$. 
 
 The local surface brightness of the Ly$\alpha$ emitting regions 
 between $z\approx2$ and $z\approx3$ is 
 consistent with each other (left panel of Figure~\ref{fig:SB_sector_corr}). 
 In this section, we further demonstrate that 
 cosmic density evolution may not be a major 
 factor of causing the evolution of the 
 circularly-averaged SB from $z\approx3$ to 
 $z\approx2$.  
 
Following the expansion of the Universe,
the cosmic mean density evolves as $(1+z)^{-3}$. 
Given the standard convention of defining a dark matter halo
as 200 $\times$ the mean density, 
the  density in the halo systematically
decreases from $z=3-2$. 
In the standard $\Lambda$CDM Universe, the cosmic scale 
factor $a(z)$ is proportional to $\frac{1}{(1+z)}$. At $z>2$, the cosmic  
density 
$\rho(z)$ is proportional to $\frac{1}{a(z)^3}$, and the density 
at $z=3.1$ ($\rho(z=3.1)$) and that at $z=2.3$ ($\rho(z=2.3)$) 
have the following relation: 
\begin{equation}
\rho(z=3.1)=  \frac{(1+2.3)^3}{(1+3.1)^3}\times \rho(z=2.3) = 1.92\times \rho(z=2.3). 
\end{equation}
The average density of a halo at $z=3.1$ is 
1.92 $\times$ the halo at $z=2.3$. 
If QSO halos follow this general 
halo density evolution, and further, if the cool gas 
density is proportional to the halo matter density, 
then the Ly$\alpha$-emitting cool 
gas density in QSO halo at $z=3.1$ is expected 
to be 1.92 $\times$ that at $z=2.3$. 
Then, the SB$_{\rm{Ly\alpha}}$ could evolve 
accordingly as the evolution of cool gas density. 

The optically thin photo-ionization model
is a favored scenario
for nebular Ly$\alpha$ emission around AGN (e.g., Heckman
et al. 1991a; Cantalupo et al. 2014; 
Arrigoni Battaia et al. 2015, 2016; Cai et al.
2017; see also discussions in Cai et al. 2018).
In the optically thin approximation, the Ly$\alpha$ surface brightness 
SB$_{\rm{Ly\alpha}}\propto n_{\rm{H}} N_{\rm{H}}$, where $n_{\rm{H}}$ is the number density of the hydrogen, 
and $N_{\rm{H}}$ is the column density which is also 
proportional to the gas number density. 
From above, we know that the QSO halo density at $z=3.1$ 
could be 1.92 $\times$ the density of halo at $z=2.3$.
If the gas in the halo is largely optically thin, 
then we expect that the $SB_{\rm{Ly\alpha}}\propto n^2$, where $n$ is the gas density. Then, we expect: 
\begin{equation}
 SB_{\rm{Ly\alpha}, z=2.3} 
= \frac{1}{3.7} \times SB_{\rm{Ly\alpha}, z=3.1},  
\end{equation}
The Equ.(8) suggests that if the dimming of the circular-averaged SB is due to the global 
evolution of the CGM density, then the characteristic redshift-dimming-corrected 
 SB$_{\rm{Ly\alpha}}$ at $z\approx3.1$ could be 
a factor of $3.7 \times$ that at $z\approx2.3$. 
This is not consistent with our current observations. 
As described in the previous section, the 
characteristic Ly$\alpha$ emission at $z\approx2$ is  consistent with that at $z\approx3$, but 
$z\approx2$ halo has a lower covering factor comparing to $z\approx3$. 
This suggests that the local densities of the cool, Ly$\alpha$-emitting gas 
have little evolution from $z\approx3$ -- $z\approx2$ as the cosmic expansion, 
as expected from cosmological theories (the CGM is not in the linear part of the growth of perturbations).

Note the above conclusion is {drawn} under the optically thin photoionization scenario. 
The optically thick photoionization is not favoured by our observations. 
If the Ly$\alpha$ nebular emission is mainly contributed by optically thick gas, 
then the strong correlation between the luminosity of Ly$\alpha$ ($L_{\rm{Ly\alpha}}$) 
and the QSO ionizing luminosities should exist.  Nevertheless, from Figure~\ref{fig:LLya_Lnu}, there are no obvious correlation between 
nebular $L_{\rm{Ly\alpha}}$ and the QSO ionizing luminosities in our sample\footnote{The ionizing luminosity of the QSO is 
determined by rescaling the composite spectrum 
from Lusso et al. (2015) to the 
QSO luminosity at the rest-frame 1350 \AA. 
If the gas is optically 
thick to Lyman continuum photons ($N_{\rm{HI}}$ $\gtrsim  10^{17.2}$ cm$^{-2}$), 
the cool gas will behave like a ‘mirror’ and 
the SB thus follows the relation $SB_{\rm{Ly\alpha}} \propto
L_{\rm{\nu LL}}$ 
(Hennawi \& Prochaska 2013). 
If the source is bright enough
to keep the gas highly ionized, i.e., the gas is optically thin ($N_{\rm{HI}} <$ 10$^{17.2}$ 
cm$^{-2}$). It can be shown that (Hennawi \& Prochaska 2013) 
the Ly$\alpha$ surface brightness is $SB_{\rm{Ly\alpha}} \propto n_{\rm{H}} N_{\rm{H}}$. Therefore, 
in the optically thin regime, the 
observed surface brightness should not depend on the luminosity
of the targeted QSOs, but on the density of cool gas.},  indicating optically thick scenario is not favoured by our observations. 

 \subsubsection{Evolution of Mass Threshold of Hot Halo}

The decreasing of the Ly$\alpha$ emitting regions from $z\approx3$ 
to $z\approx2$ may also be interpreted using the mechanisms of the cool 
gas penetration in the massive halos \citep{dekel09}. 
Dekel et al. (2009) study the penetration of 
cold gas into the halo as a function of halo mass 
and redshift. They find that, from the simulation, halos at $z\gtrsim3$ with 
$M>10^{13}\ M_\odot$ are dominated by hot gas, but at 
$z\approx2$, the mass threshold of hot halo decreases to $M\approx10^{12}$ M$_\odot$. 
This indicates that QSO halos with a typical halo mass of 
$M\sim 10^{12.5}\ M_\odot$ 
(e.g., White et al. 2012) is more likely to be dominated by hot gas at $z\approx2$ while significant 
amount of cool gas 
could still penetrate such halos at $z\approx3$. This effect may  
reduce the surface brightness of Ly$\alpha$ emission from $z\approx3$ to 
$z\approx2$. 
Note that the cool gas penetration scenario would 
hold for any Ly$\alpha$ powering mechanisms 
 (see \citet{arrigoni19}), i.e., the decreasing 
of the cool gas penetration in massive halos may yield 
a decrease of the Ly$\alpha$ emission. 
More quantitative analysis on whether this effect 
can decrease the covering factor of cool gas should  
be further studied by simulations, combined with
deeper observations on other transition lines.


 \subsection{Ly$\alpha$ morphology and its Evolution 
 from $z\gtrsim3$ to $z\approx2$}

\subsubsection{Summary of the Ly$\alpha$ Morphology at $z\approx2$ in our KCWI sample}

 From Borisova et al. (2016) and Arrigoni Battaia et al. (2019), the majority of the 
 MUSE radio-quiet nebulae at $z\gtrsim 3$ have symmetric morphology. 
 In particular, nebulae with scales smaller than 100 kpc 
 have circular morphologies. Nevertheless, our new KCWI observations suggest that nebulae 
 at $z\approx2$ have  more irregular 
 and asymmetric morphologies. In the following, we briefly comment on the morphology, 
 kinematics and surface brightness of  individual nebula. 
 
 From Figure~\ref{fig:optimal_extraction}, we observe that Q2121 has two spatially distinct 
 components, with a projected separation of $\approx 85$ kpc. 
 From Figure~\ref{fig:2Dspec}, the two components have similar 
 velocity and the entire Ly$\alpha$ emission has a FWHM of 
 $\approx 500$ km s$^{-1}$. Q0048 and Q1426 both have Ly$\alpha$ emitters 
 in the KCWI field of view, at the same redshift with the QSOs, 
 suggesting strongly overdense environment in both fields. 
 The bright Ly$\alpha$ emitter in  the Q0048 field is 5$''$ west 
 of the QSO center. Seven nebulae: Q2121; Q2123; 
 Q1227; Q1228; Q1230; Q1416 have projected scales of $>150$ kpc and 
 therefore extend beyond the FoV of the KCWI medium slicer. Their spatial extents 
 are more similar 
 as enormous Ly$\alpha$ nebulae (e.g., Cai et al. 2017; Arrigoni Battaia et al. 2018) 
 with the projected size of $\gtrsim200$ kpc. 
 Q2127, Q0814, Q0107, Q2125 and Q0848 are 
 among the lowest Ly$\alpha$ surface brightness in the sample whose projected 
 scales are $<100$ kpc. 
 Our KCWI observations 
 suggest that nebulae around QSOs at $z\approx2$ may be generally more 
 asymmetric, irregular, and having stronger field-to-field variations than that at $z\approx3$. 
 
 \subsubsection{Evolution of the Ly$\alpha$ Morphology}
 
The morphology of Ly$\alpha$ nebulae contains 
information about the cool gas 
distribution and geometry in the halos. 
 Morphology can also be used to directly compare with cosmological simulations (e.g., Weidlinger et al. 2005; 
 Cantalupo et al. 2014). In \S3.4, we quantified the asymmetry of the Ly$\alpha$ light distribution following Arrigoni Battaia et al. (2019).  
 From \S3.4, we found that the median 
ratio between the 
semiminor axis and semimajor axis at $z\approx2$ is 
$\alpha_{\rm{z=2}}=0.54$, with a scatter of 0.19. For a comparison, at $z\approx3$, 
Arrigoni Battaia et al. (2019) measured the average ratio of  $<\alpha_{\rm{z=3}}>=0.71$, with a scatter of 0.13 (also see Figure~\ref{fig:alpha_size}). 
This tentitatively suggests that nebulae at $z\approx2$ may be more asymmetric or clumpy than that at $z\approx3$. Such an asymmetry 
could further contribute to the lower average surface brightness at $z\approx2$. 
{ This can be further seen from Figure~\ref{fig:2Dspec} which the 2D spectra of the nebulae at
 $z\sim3$ (Borisova) look smooth in the velocity space, while nebulae in the 
current survey at $z\sim2$ in our sample are more clumpy.}


Galaxy structure is a powerful method for determining 
whether a galaxy is undergoing a recent major merger.  
Lotz et al. (2008, 2010) find that the asymmetry 
is sensitive to mergers with mass ratios of 1:4 or less. 
 The halo morphological merger fraction increases from 
 $z\approx3- 2$ and decreases from $z=2-1$. 
 The merger fraction 
 is highest around $z\approx2$ (e.g., Conselice et al. 2008). 
 Lopez-Sanjuan et al. (2009) further point out that the merger 
 rate at $z\approx 2$ could be a factor of $2\times$ that at $z\gtrsim3$. 
 Such an increased merger fraction at $z\approx2$ may result in 
 the asymmetric gas morphology we observe at $z\approx2$.

\subsection{Offsets between the systemic redshift and Ly$\alpha$ redshift}

{ As indicated in \S4.1, we  
consider three cases: (1) systemic redshift determined from SDSS pipeline ($z_{\rm{pipe}}$); 
(2) systemic redshift determined by \mgii\  emission ($z_{\rm{MgII}}$), correcting 
the luminosity-dependent small and known offset between \mgii\ and 
systemic redshift (Richards et al. 
2002; Shen et al. 2016); and (3) systemic redshift determined by the principal component analysis, with the 
reference sample has been chosen to have an automated redshift corresponding to the location of the maximum 
of the \mgii\ emission line ($z_{\rm{PCA}}$). These three redshifts are all drawn from the SDSS DR12 QSO catalog 
\citep{paris17}. }

The uncertainties of these redshifts 
 are known to be $\delta_z \sim 0.003$ ($\approx300$ km s$^{-1}$) 
(also see Arrigoni Battaia et al. 2016).
 In Figure~\ref{fig:2Dspec}, we show the systemic redshift determined by SDSS pipeline ($z_{\rm{pipe}}$) (purple 
 vertical line) relative to the center of the integrated nebular Ly$\alpha$ emission (marked as zero in the x-axis). 
 We further show the center of QSO Ly$\alpha$ emission using the orange vertical line.  
 We calculate the velocity shift ($v_{\rm{off}}$) between the \mgii-derived systemic redshift ($z_{\rm{sys}}$) and 
 the nebular Ly$\alpha$ redshift ($z_{\rm{Neb, Ly\alpha}}$), and the expression is as follows: 
 \begin{equation}
 v_{\rm{off}}= (z_{\rm{sys}}- z_{\rm{Neb, Ly\alpha}})/(1+z_{\rm{Neb, Ly\alpha}})\times c
 \end{equation}
 The velocity shifts range from -3947 to +813 km s$^{-1}$, with a median velocity shift of -1081 km s$^{-1}$. 
 In our KCWI sample, only Q1416 and Q2125 have the $z_{\rm{sys}}$ 
 redshifted than the nebular Ly$\alpha$ redshift ($z_{\rm{Ly\alpha}}$). All other QSOs ($87\%$ of our sample) 
 have $z_{\rm{sys}}$ blueshifted than the nebular Ly$\alpha$ redshift. 
 Similar results have already been reported in \citet{arrigoni19} at $z\approx3$:  
 at $z\approx3$, \citet{arrigoni19} show that the  velocity shift between the
 QSO systemics and the nebular Ly$\alpha$ redshift range 
 -6000 km s$^{-1}\le $ $v_{\rm{offset}} \le 2000$ km s$^{-1}$, and 
80\% nebulosities have negative velocity
shifts, with the median value of  
$\Delta v_{\rm{median}} = -782$ km s$^{-1}$. Similar results and discussions can 
also be found in Arrigoni Battaia et al. (2019). 
 Note that if we use $z_{\rm{PCA}}$ as the systemic redshift, then the velocity offset ranges from 
 -4281 km s$^{-1}$ -- 1463 km s$^{-1}$, with the median velocity offset of -553 km s$^{-1}$, with 
 the 1-$\sigma$ scattering of 1417 km s$^{-1}$. 
If we use  $z_{\rm{MgII}}$ as the systemic redshift, the velocity shifts range from 
-1586 km s$^{-1}$ to $1646$ km s$^{-1}$, with the median offset of -92 km s$^{-1}$, 
with the 1-$\sigma$ scattering of 877 km s$^{-1}$.

It is surprising to note that most of the SDSS QSO systemic 
redshifts in our sample are 
blueshifted compared to 
the nebular Ly$\alpha$ emission, {and also, most SDSS QSO systemic 
redshift has large offset with respect to the nebular Ly$\alpha$ redshift. 
Currently, we still do not 
find a good explanation for such a systematic shift.} 
We only can conclude that with an intrinsic
uncertainty of 300 km s$^{-1}$
for the empirical calibration of the emission (e.g., Shen et al. 2016), such 
systemic redshift estimates may not be optimal for 
our ultraluminous QSO sample. 
Future efforts are definitely needed to better constrain
this fundamental parameter. 
{A better strategy to obtain a more accurate systemic redshift (down to a few tens of km s$^{-1}$)  may require sub-mm observational campaign in future.}
 These sub-mm 
observations can help us to evaluate whether the systemic 
redshift derived from \mgii\ emission is accurate or not 
for such a high luminosity QSO sample at $z\approx2$.

\section{Summary}

In this paper, we conduct a systematic, blind survey of 
Ly$\alpha$ nebulae around ultraluminous Type-I QSOs at $z\approx2$ 
using the Keck Cosmic Web Imager (KCWI). 
  {This survey allows us to directly compare with similar integral-field-spectroscopic studies at 
$z\gtrsim3$ using MUSE and to study the evolution of the cool gas in massive halos.}
The main conclusion of this paper are as follows:

(1)  We find that {14 out of 16 QSOs}  at $z\approx2$ are 
associated with Ly$\alpha$ nebulae with projected linear sizes larger than 
50 physical kpc (pkpc). Among them, four nebulae 
have large Ly$\alpha$ emission with surface brightness of $SB_{\rm{Ly\alpha}} >10^{-17}$ 
erg s$^{-1}$ cm$^{-2}$ arcsec$^{-2}$ on $>100$ kpc, and their scales extend beyond the KCWI 
field of view.

(2) Our KCWI results suggest that the nebulae at $z\approx2$ are 
one order of magnitude brighter and more extended than 
previous narrowband surveys indicate (e.g., Arrigoni Battaia et al. 2016). 
{ We suggest that this is partially due to the limitation of the narrowband survey technique 
and a significant fraction of the diffuse Ly$\alpha$ emission may have be missed in the narrow-band 
imaging because of large offsets between the true and the estimated quasar systemic redshift from 
the MgII line. (see \S4.1). }
The circularly-averaged Ly$\alpha$ profile is also much more brighter than that 
of Lyman break galaxies (LBGs) (e.g., Steidel et al. 2011) and Ly$\alpha$
emitters \cite[e.g.,][]{wisotzki16}. No regular rotational kinematic 
patterns have been found in the diffuse Ly$\alpha$ emission for our 
QSO sample at $z\approx2$.

(3) We directly measure the circularly-averaged surface 
brightness (SB) at $z\approx2$ using the integral field spectroscopy, and we 
perform a direct comparison with the $z\approx3$ results using VLT/MUSE \citep{borisova16, arrigoni19}. 
 The typical circularly-averaged SB profile of Ly$\alpha$ nebulae around $z\approx2$ 
 QSOs can be described by a power-law with the slope of -1.8, same as the 
 one measured for the nebulae around $z\approx3$ QSOs, however its normalisation 
 is about a factor 0.4 dex fainter than the SB profiles at $z\approx3$, after correcting for 
 the different redshift dimming.  A larger FoV could further constrain the profile on a larger 
 radii, especially to differentiate the exponential and power-law profiles \cite[e.g.,][]{arrigoni19}. 


(4) Our analysis suggests that 
the Ly$\alpha$ emitting cool gas in the QSO halos at $z\approx2$ may have smaller 
cover fraction (see \S4.2). This may be one of the reasons for the lower 
circularly average SB from $z\approx3$ to $z\approx2$. 
Further, nebulae around $z\approx2$ QSOs appear to have more irregular 
and asymmetric morphologies compared to QSO nebulae at 
$z\approx3$ as quantified by the ratio between the 
semiminor and semimajor axis above the 2$\sigma$ 
detection level (and denoted by $\alpha$ as in 
Arrigoni Battaia et al. 2019).  In particular, 
the average $\alpha$ at $z\approx2$ is $\sim40$\% 
smaller than that at $z\approx3$. 
Taking into account these different morphologies, the 
local SB values of the Ly$\alpha$ region, once corrected for redshift-dimming,  
become similar between the different redshifts, 
especially at the radii $>60$ kpc (see e.g., Figure~\ref{fig:SB_sector_corr}). . 

Taken all together, our KCWI results suggest that 
from $z\approx 3$ to $z\approx 2$, the covering factor
and possibly the overall mass of cool Ly$\alpha$ emitting CGM 
could decrease, assuming the same opening angle at $z\approx2$ and 
$z\approx3$. For the Ly$\alpha$ emitting region, 
 the typical densities of the cool gas around QSOs do not strongly evolve from 
$z\approx3$ to $z\approx2$. Last but not least, we note that there is still 
a large scatter and variation in the Ly$\alpha$ surface brightness  
and covering factor among individual 
systems and that a larger sample would be necessary in order 
to obtain a better statistical 
analysis and comparison among 
different redshifts and QSO properties. 


\vspace{0.2in}

{\bf Acknowledgement:} 
We acknowledge the discussion of the data reduction and analysis 
with Chris Martin, Matt Matuszewski, James D. Neill, Erika Hamden, and Donal Sullivan. 
ZC acknowledges the supports provided by NASA through
the Hubble Fellowship grant HST-HF2-51370 awarded by the
Space Telescope Science Institute, which is operated by the
Association of Universities for Research in Astronomy, Inc., for
NASA, under contract NAS 5-26555. 
SC gratefully acknowledges support from 
Swiss National Science Foundation grant PP00P2\_163824.


\newpage
\vspace{5.0in}
\begin{sidewaystable*}[!h]
\caption{QSO Properties at $z\approx2$ in our QSO snapshot program}
\label{table:QSO_snapshot}	
\centering 
\begin{tabular}{| c | c | c | c | c | c |c|c| c| c | c| c| c| c | c | } 
\hline\hline 
Name & RA & DEC  &  $z_{\rm{Ly\alpha}}$$^{(a)}$  &$z_{\rm{pipe}}$ & $z_{\rm{PCA}}$  & $z_{\rm{MgII}}$ & $i$-mag$^{(b)}$  & Log$_{\rm{10}}L_{\nu{_{\rm{LL}}}}^{(c)}$  & $L_{\rm{Ly\alpha}}$  & FWHM$_{\rm{Ly\alpha}}$ & Projected   &  $2\sigma_{\rm{SB}}^{(d)}$  & Seeing & Asymmetry \\ 

 &   &    &    &  & &   &   &  & (10$^{43}$ erg s$^{-1}$) & (km s$^{-1}$)  & Extent  (kpc) & {cgs$^{(\rm{g})}$}  &  (arcsec)& $\alpha$ \\ 
\hline
\hline
Q0048+0056  & 00:48:56.34  &  +00:56:48.1  &  2.328 &  2.327 & 2.323 & 2.327 & 17.99$\pm0.01$ & 31.20   & 2.28  & $542\pm28$ & 104& 1.8 & 0.9$''$ & 0.50 \\
\hline
Q0050+0051  & 00:50:21.22  &  +00:51:35.0  &  2.241 &  2.220 & 2.219 & 2.222 & 17.82$\pm0.01$ &  31.22 & 2.00 & $481\pm29$ & 116 & 1.8  & $0.9''$ & 0.59 \\
\hline
Q0052+0140 &  00:52:33.67 &   +01:40:40.8  &  2.309  &  2.291 & 2.302 & 2.300 & 17.34$\pm0.01$  &  31.45 & 2.03 & $549\pm29$ & $127$ &1.7 & $0.8''$ & 0.61  \\
\hline 
Q0107+0314 &  01:07:36.90 & +03:14:59.2 &    2.280  & 2.269 & 2.264 &  2.262& 18.01$\pm0.01$ &  31.17 & 1.52 & $410\pm30$  & 114&1.8  & $0.9''$ & 0.51 \\ 
\hline 
Q0814+3250 &  08:14:01.38 &  +32:50:48.1   & 2.219  &  2.211 & 2.223& 2.222 & 18.50$\pm0.01$ &  30.94  &  0.15  & $662\pm32$ & 85& 1.7 & $0.8''$ & 0.31 \\ 
\hline
Q0848-0114$^{(e)}$ &  08:48:56.95 & -01:14:58.9  & 2.300   & 2.293 &  2.301 & 2.302  & 18.40$\pm0.01$ & 30.91   & 0.13  & $378\pm21$ & 28 &2.3 & $1.1''$ & 0.50 \\
\hline 
Q1227+2848  & 12:27:27.48 & +28:48:47.9  &   2.265  & 2.255 & 2.266  & 2.268 &  17.73$\pm0.01$  &  31.27  & 5.77 & $557\pm31$ &  $>164$ & 1.8 & $0.9''$ & 0.36 \\ 
\hline
Q1228+3128 &   12:28:24.97 & +31:28:37.7  & 2.214    & 2.220 & 2.199&  2.231 & 15.68$\pm0.01$ & 32.07 & 12.29 & $1175\pm39$ & $>124$ & 1.9  &  $0.9''$ & 0.52\\ 
\hline 
Q1230+3320 &  12:30:35.47  & +33:20:00.5  &   2.323  &  2.314 & 2.309& 2.313& 17.76$\pm0.01$ & 31.29  & 12.40 & $643\pm22$ & $>204$& 2.0  & $1.0''$ & 0.73\\ 
\hline 
Q1416+2649 &  14:16:17.38 & +26:49:06.2  & 2.293   & 2.296  & 2.340 &2.301 & 18.04$\pm0.01$  &  31.16 & 5.00 & $829\pm32$ & $>141$& 2.1   & $1.1''$ & 0.52\\ 
\hline 
Q1426+2555  & 14:26:35.86 &  +25:55:23.7 & 2.255  &  2.248 & 2.249 & 2.256 &  17.60$\pm0.01$ &  31.32 & 3.88 & $622\pm26$ &  96  & 1.9 &  $0.9''$ & 0.84\\ 
\hline
Q1444+3904 & 14:44:55.89 & +39:04:00.7 & 2.250 & 2.250 & 2.251 & 2.250 & 18.40$\pm0.01$ & 31.26 & 10.09 & $951\pm40$ & $101$ & 2.1  & $1.1''$ & 0.70\\
\hline
Q2121+0052 & 21:21:59.04  &  +00:52:24.1 &  2.373 & 2.367   & 2.372 & 2.377 & 18.07$\pm0.01$ &  31.19 & 4.56 & $656\pm32$ & $>141$ & 1.8 & $0.8''$ & 0.38\\
\hline 
Q2123-0050 & 21:23:29.46   &  -00:50:52.9  &  2.280 & 2.250 & 2.266 & 2.271 &  16.34$\pm0.01$  & 31.84 & 3.59 &  $812\pm35$  & $>154$ & 2.0 & $1.0''$ & 0.50\\
 \hline
Q2125+0112$^{(f)}$ & 21:25:11.83 &  +01:12:22.32 & 2.199 &  2.200 & 2.202 & 2.203 & 18.51$\pm0.01$ & 31.01 & 1.29 & $514\pm60$ & 19 & 2.3  &  $1.1''$ & 0.60\\ 
\hline
Q2127+0049 & 21:27:47.43  &  +00:49:29.5  &  2.269 & 2.245  & 2.251 & 2.261& 17.53$\pm0.01$ &   31.35 &  1.10  & $820\pm41$  & 58 &1.8 &  $0.9''$ & 0.69\\
\hline
\end{tabular}
{\footnotesize{
\leftline {(a): The Ly$\alpha$ emission has an extremely high SNR, yielding the typical uncertainties of the Ly$\alpha$ line center of 0.02 \AA,}
\leftline{  corresponding to the statistical error of Ly$\alpha$ redshift of $\lesssim0.001$.} }}
\leftline {(b): The $i$-magnitude error of these sources have a typical error of $i$-magnitude $\le 0.01$. The magnitude 
are taken from the SDSS Data Release 12. }
\leftline {(c): The ionizing luminosity of the QSO is 
determined by rescaling the composite spectrum 
from Lusso et al. (2015) to the 
QSO luminosity }
\leftline {at the rest-frame 1350 \AA. The typical statistical error is $\le 1$\% level.} 
\leftline {(d): The surface brightness uncertainty is calculated by integrating over 1 \AA\ at around $\lambda \approx 1230$\AA, between Ly$\alpha$ and NV emission, using circles with 1$''$ diameter. } 
\leftline {(e): Q0848-0114 was taken under a partially thin cloudy condition.} 
\leftline {(f): Q2125+0112 was taken under a partially thin cloudy condition.} 
\leftline {(g): cgs: 10$^{-18}$ erg s$^{-1}$ cm$^{-2}$ arcsec$^{-2}$} 
\vspace{3.0in}
\end{sidewaystable*}


\end{document}